\documentclass[prb,aps,showpacs,twocolumn,floatfix]{revtex4-1}
\usepackage{amsmath, amsfonts, amssymb, bm, graphicx}

\newcommand{\bS}{{\bm S}}
\newcommand{\bk}{{\bm k}}
\newcommand{\bp}{{\bm p}}
\newcommand{\bpp}{{{\bm p}_\perp}}
\newcommand{\spp}{{p_\perp}}
\newcommand{\sppp}{{p^\prime_\perp}}
\newcommand{\bppp}{{{\bm p}^\prime_\perp}}
\newcommand{\bsig}{{\bm\sigma}}
\newcommand{\balpha}{{\bm\alpha}}
\newcommand{\e}{{\rm e}}
\newcommand{\ii}{{\rm i}}
\newcommand{\ad}{a^\dag}
\newcommand{\cd}{c^\dag}
\newcommand{\fd}{f^\dag}

\newcommand{\ua}{{\uparrow}}
\newcommand{\da}{{\downarrow}}
\newcommand{\vac}{\vert0\rangle}
\newcommand{\FS}{\vert{\rm FS}\rangle}

\begin{document}

\title{Tunable unconventional Kondo effect on topological insulator surfaces}

\author{L. Isaev$^{1, 2}$}
\author{G. Ortiz$^3$}
\author{I. Vekhter$^2$}
\affiliation{
 $^1$JILA, NIST \& the University of Colorado, Boulder, CO 80309, USA \\
 $^2$Department of Physics and Astronomy, Louisiana State University, Baton
 Rouge LA 70803, USA \\
 $^3$Department of Physics and Center for Exploration of Energy and Matter,
 Indiana University, Bloomington IN 47405, USA
}

\begin{abstract}

 We study Kondo physics of a spin-$\frac{1}{2}$ impurity in electronic matter
 with strong spin-orbit interaction, which can be realized by depositing
 magnetic adatoms on the surface of a three-dimensional topological insulator.
 We show that magnetic properties of topological surface states and the very
 existence of Kondo screening strongly depend on details of the bulk material,
 and specifics of surface preparation encoded in time-reversal preserving
 boundary conditions for electronic wavefunctions.
 When this tunable Kondo effect occurs, the impurity spin is screened by purely
 orbital motion of surface electrons.
 This mechanism gives rise to a transverse magnetic response of the surface
 metal, and spin textures that can be used to experimentally probe signatures
 of a Kondo resonance.
 Our predictions are particularly relevant for STM measurements in ${\rm Pb
 Te}$-class crystalline topological insulators, but we also discuss
 implications for other classes of topological materials.

\end{abstract}
\maketitle

\section{Introduction}
\label{sec_intro}

Recent explosion of interest in topological insulators (TIs)
\cite{zhang-2009-1,hasan-2010-1,qi-2011-1} is due in large part to the fact
that they support metallic states on their surfaces.
The existence of these states (and hence the metallicity) results from the
non-trivial topological nature of the Bloch wavefunctions in the conduction and
valence bands of the bulk material, and is a robust feature.
In contrast, the quantum numbers associated with those surface states are not
determined by topology alone.
Therefore they may vary from material to material, and depend on the surface
preparation.
Understanding physical consequences of this non-universal behavior is one
of the foci of our paper.

A typical cartoon picture of surface states in a TI consists of
spin-momentum-locked energy branches of a massless Dirac spectrum.
This description cannot be universally accurate.
A crystal boundary breaks the inversion symmetry and gives rise to
strong, rapidly varying in space, electric fields that define an effective
surface potential for the electrons.
Interplay between these field gradients and the bulk inter-atomic spin-orbit
interaction (SOI), responsible for the non-trivial topological aspects of these
materials, renders this potential momentum and spin dependent.
As we show below, this ensures that measurable properties of the surface states
cannot be determined by topological arguments alone.
Details associated with a crystal surface can be accounted for via effective
boundary conditions (BCs) for the electron wavefunctions
\cite{volkov-1981-1,satanin-1984-1}, and are completely excluded from the
topological arguments involving only the bulk band structure.
In this context, Ref. \onlinecite{isaev-2011-1} argued that appropriate BCs are
essential for a sensible formulation of a bulk-boundary correspondence in TIs.
Moreover, Ref. \onlinecite{zhang-2012-1} pointed out a dependence of the spin
texture of surface Dirac cones on the crystallographic orientation of the
surface, even for simple BCs.

In this paper we show that the spin behavior of surface states in
three-dimensional (3D) TIs is highly sensitive to both the bulk band structure
and surface properties.
We consider semiconductors with different crystal symmetry: cubic lead
chalcogenides (${\rm Pb Te}$ or ${\rm Pb Se}$) and tetragonal ${\rm Bi_2
Se_3}$-like TIs, and demonstrate that magnetic probes (such as an external
field or quantum impurities) can be used to efficiently differentiate between
these two classes.
Crucially, the sensitivity of TI surface states to the surface manipulation
allows one to use TIs as a controllable environment for studying spin-dependent
correlated phenomena in the presence  of strong SOI.

While some of the unusual magnetic phenomena that we argue for can be probed by
measuring the response to a uniform magnetic field, in this paper we focus on
the physics of a spin-$\frac{1}{2}$ impurity deposited on the surface of a
3D TI.
Kondo screening, whereby the impurity spin at low temperatures forms a singlet
state with the Fermi sea, is one of the earliest and lucid examples of
correlated many-body physics \cite{wilson-1975-1} that remains relevant in
contexts ranging from heavy fermion systems \cite{hewson-1997-1,cox-1999-1} to
nanoscience \cite{kikoin-2011-1}.
Advances in scanning tunneling microscopy (STM) allowed observation of this
phenomenon on the atomic scale \cite{madhavan-1998-1,li-1998-1,pruser-2011-1},
and granted access to manipulation of individual Kondo resonances
\cite{tsukahara-2011-1}.
Testing surface states of TIs via STM \cite{roushan-2009-1} complements
spin-polarized angle-resolved photoemission (ARPES) measurements
\cite{hsieh-2009-1,herdt-2013-1}, and gives a direct probe of the Kondo effect.

In its simplest form, Kondo screening involves only spin degrees of freedom of
the conduction electrons.
Hence it is sensitive to the spin-$SU (2)$ symmetry breaking, provided in our
case by the SOI.
Previous works have shown that the Kondo effect survives in the presence of
spin-orbit scattering \cite{bergmann-1986-1,wei-1989-1,meir-1994-1}, and weak
(compared to the bandwidth) Rashba or Dresselhaus band SOI
\cite{malecki-2007-1,yanagisawa-2012-1,mastrogiuseppe-2014-1,zitko-2011-1,
feng-2010-1,hu-2013-1}.
In some cases, the latter can actually enhance the Kondo resonance
\cite{isaev-2012-1,zarea-2012-1}.
The {\it strong} SOI regime is even more intriguing.
Indeed, the SOI can be viewed as a momentum-space magnetic ``field'' that
aligns electron spins along a particular direction (e.g. perpendicular to its
momentum).
When this field is large enough the spin degree of freedom of conduction
electrons is effectively lost and cannot participate in the spin-flip
scattering leading to the Kondo effect.
Nevertheless there is substantial theoretical evidence \cite{zitko-2010-1,
tran-2010-1,feng-2010-1,mitchell-2013-1,orignac-2013-1,xin-2013-1} indicating
that a magnetic impurity on a TI surface is screened by the surface metal.
Remarkably the physical nature of this effect and spatial structure of the
screening states have never been elucidated in the context of TIs.
Understanding this phenomenon is also important from an experimental
perspective because magnetic probes (e.g. impurities or magnetic field) coupled
to surface electrons can be used to differentiate trivial and topologically
non-trivial matter, providing an alternative to ARPES-based techniques.

We demonstrate that the strong SOI leads to an unconventional Kondo effect with
an impurity spin screened by purely {\it orbital motion} of surface electrons.
Specifically, we consider a simple band model of a 3D  TI, and derive an
effective Kondo Hamiltonian that governs the dynamics of the impurity spin at
the TI boundary (that does not break time-reversal symmetry), taking into
account the {\it full 3D} spatial dependence of surface-state wavefunctions.
Because of the SOI this Kondo exchange has an $XXZ$ structure and, in general,
is strongly anisotropic.
At low energies, the impurity spin forms a singlet state with the total
electron angular momentum, and the system exhibits an emergent $SU (2)$
symmetry, which is responsible for the Kondo resonance.
The SOI also gives rise to a transverse magnetic response when an external
magnetic field applied normal to the surface results in an in-plane electron
spin polarization, which may lead to interesting magneto-electric phenomena
under driving fields.
This response is significantly stronger than an analogous effect on metallic
surfaces with Rashba SOI \cite{lounis-2012-1,chirla-2013-1}.

In Sec. \ref{sec_model}, we describe our minimal model of a 3D TI and calculate
its surface spectrum.
Here we consider a continuum version of a lattice model studied in Ref.
\onlinecite{isaev-2011-1}.
Emphasis is put on the physical meaning of the quantum numbers involved in the
effective description of electronic states.
In Sec. \ref{sec_ss-mag_int} we explain how both surface and bulk states play a
role in determining the specific mathematical form of the relevant operators
involved in the effective coupling between surface electrons of the TI and the
magnetic impurity.
Here, we contrast ${\rm Bi_2 Se_3}$ and ${\rm Pb Te}$-class materials.
Section \ref{sec_eff-surface-H} establishes the effective $XXZ$ Kondo
Hamiltonian that governs coupling of these surface states to magnetic
impurities, and explains why this is a single-channel Kondo Hamiltonian despite
its apparent two-channel form.
In our approach we control the surface properties through BCs for electronic
wavefunctions \cite{isaev-2011-1} and show that surface manipulation provides
an effective way of tuning parameters in the effective low-energy Kondo model
and can be used to completely suppress the spin-flip terms and destabilize the
Kondo effect.
We study the physical properties of the effective model and its
unconventional Kondo physics in Sec. \ref{sec_results}.
In particular, we investigate the transverse magnetic response to an external
magnetic field and point to the resulting transverse spin textures as a
distinctive characteristic of the Kondo screening cloud in strong SOI
materials.
We also show how one can tune the Kondo effect and the characteristic
temperature $T_K$ via surface manipulation.
Our results can be directly verified in STM measurements in crystalline TIs
like the lead-tin solid alloys ${\rm Pb_{1 - x} Sn_x Te}$, but the above
unconventional Kondo physics can also be observed in well-studied ${\rm Bi_2
Se_3}$ and ${\rm Bi Sb}$.
Finally, Sec. \ref{sec_discussion} provides a summary and an outlook with
questions that still remain open.
Two appendices with technical derivations complete the paper: Appendix
\ref{sec_app-a} addresses the very important problem of self-adjoint extensions
of unbounded Hermitian operators, of key relevance to the analysis of bound
surface states.
Appendix \ref{sec_app-b} exploits the axial symmetry of the problem to
construct surface states with well-defined total angular momentum.

\section{Simple continuum model for topological insulators}
\label{sec_model}

\subsection{Model Hamiltonian}
\label{sec_model_subsec_model}

To describe electronic states in a TI we use Dimmock's model
\cite{carter-1971-1,kang-1997-1}, defined by the modified Dirac Hamiltonian
\begin{equation}
 H_{\rm D} = v (\balpha \cdot {\bm p}) + \beta \biggl( \Delta +
 \frac{{\bm p}^2}{2 m^*} \biggr).
 \label{eq:dimmock-H}
\end{equation}
This effective Hamiltonian involves two spinful bands (conduction and valence)
of opposite parity separated by an energy gap $2 \Delta$ and is written in
terms of the $4 \times 4$ Dirac matrices
\begin{displaymath}
 \balpha = (\sigma^x \otimes \bsig) =
 \begin{pmatrix}
  0 & \bsig \\
  \bsig & 0
 \end{pmatrix}, \quad
 \beta = (\sigma^z \otimes {\bm 1}) =
 \begin{pmatrix}
  {\bm 1} & 0 \\
  0 & -{\bm 1}
 \end{pmatrix},
\end{displaymath}
with $\bsig = (\sigma^x, \sigma^y, \sigma^z)$ denoting the usual Pauli
matrices, and ${\bm 1}$ -- the unit $2 \times 2$ matrix.
In Eq. \eqref{eq:dimmock-H} the effective mass $m^*$ accounts for contributions
from remote bands, and the velocity scale $v$ is proportional to the momentum
matrix element between conduction and valence Bloch states.
In the following we shall adopt units with $\hbar = 1$.

The Dimmock Hamiltonian \eqref{eq:dimmock-H} provides a standard description of
electronic spectra in lead chalcogenides near one of the 8 equivalent
$L$-points in the Brillouin zone.
Note that despite the presence of SOI Eq. \eqref{eq:dimmock-H}, is written in
the basis of direct-product states \cite{kang-1997-1} $\vert L_6^\pm \rangle
\otimes \vert \sigma \rangle$, where $L_6^\pm$ denote spinor one-dimensional
representations of $D_{3 d}$ corresponding to the conduction ($L_6^-$) and
valence ($L_6^+$) bands, superscripts $\pm$ indicate spatial parity of the
state, and $\sigma = \ua$, $\da$ is the electron spin quantum number.
Eigenstates of $H_{\rm D}$ are 4-component envelope functions
\begin{displaymath}
 \psi ({\bm x}) =
 \begin{pmatrix}
  \psi_{c, 1} ({\bm x}) \\
  \psi_{c, 2} ({\bm x}) \\
  \psi_{v, 1} ({\bm x}) \\
  \psi_{v, 2} ({\bm x})
 \end{pmatrix}
 \!\! , \,\,
 \psi^\dag ({\bm x}) = \bigl( \psi_{c, 1}^* \, \psi_{c, 2}^* \,
 \psi_{v, 1}^* \, \psi_{v, 2}^* \bigr),
\end{displaymath}
which define the full electron wavefunction in the crystal:
$\langle{\bm x} \vert \Psi \rangle = \sum_{i = 1, 2} [\psi_{c, i} ({\bm x})
\langle {\bm x} \vert u_{c, i} (\bk_0) \rangle + \psi_{v, i} ({\bm x}) \langle
{\bm x} \vert u_{v, i} (\bk_0) \rangle] \e^{\ii \bk_0 {\bm x}}$ where
$\langle {\bm x} \vert u_{(c, v), i} (\bk_0) \rangle \e^{\ii \bk_0 {\bm x}}$
are Bloch states corresponding to band extrema at the point $\bk_0$ in the
Brillouin zone ($\bk \cdot \bp$ method).
The state $\vert \Psi \rangle$ does not need to have a definite spin quantum
number due to the SOI usually present in TIs.
In general, the indices $i = 1, 2$ describe {\it pseudospin} states whose
relation to the true spin will depend on the material.
For instance, in ${\rm Pb Se}$-like systems the gap at the $L$-point is formed
by non-degenerate representations of the single group $D_{3 d}$.
The SOI does not affect these states besides shifting their energy, so the
pseudospin states $i = 1, 2$ can be identified with the eigenstates of
$\sigma^z$ [i.e. $\vert \sigma \rangle = \vert \ua \rangle$ or $\vert \da
\rangle$] \cite{kang-1997-1}.
The (periodic part of the) Bloch basis functions can be taken as direct
products of orbital and spin parts $\vert u_{c,i} (\bk_0) \rangle = \vert u_c
(\bk_0) \rangle \otimes \vert \sigma \rangle$.

The Hamiltonian \eqref{eq:dimmock-H} has a number of conserved
``tensor'' spin operators \cite{bagrov-1990-1}. For us the important one is
\begin{equation}
 {\bm T} = \beta [{\bm\Sigma} \times \bp] =
 {\rm diag} \bigl \lbrace [\bsig \times \bp], \, -[\bsig \times \bp] \bigr
 \rbrace,
 \label{eq:tensor-spin-T}
\end{equation}
with ${\bm\Sigma}=(1\otimes\bsig)$. One can easily check that $[H_{\rm D}, {\bm
T}] = 0$. Here we will only need $T^z = \ii \beta \alpha^z (\balpha^\perp \cdot
\bpp)$, where ``$\perp$'' denotes $xy$ vector components. The Dimmock
Hamiltonian \eqref{eq:dimmock-H} can be rewritten as
\begin{equation}
 H_{\rm D} = v \bigl( \alpha^z p_z + \ii \beta \alpha^z T^z \bigr) + \beta
 \biggl( \Delta + \frac{{\bm p}^2}{2 m^*} \biggr).
 \label{eq:dimmock-H-2x2}
\end{equation}
Note, that $T^z$ is a block-diagonal matrix whose elements are proportional
to a ``Rashba'' SOI term $[\bsig \times \bpp]_z$.

Effective mass models similar to \eqref{eq:dimmock-H} emerge in many narrow-gap
semiconductors with strong SOI \cite{bir-1974-1}, such as ${\rm Bi_2 Se_3}$.
The relevant point in the Brillouin zone and the interpretation of the quantum
numbers may differ depending on the material.
For example, in ${\rm Bi_2 Se_3}$ (symmetry $D_{3 d}$ at the $\Gamma$-point)
the SOI is essential in determining gap-forming states \cite{zhang-2009-1},
hence the basis states are no longer direct products.
Even though there are still four states in the vicinity of the gap and the
effective mass description is given by Eq. \eqref{eq:dimmock-H}, the
identification of the pseudospin with real spin (as for ${\rm PbSe}$) is no
longer possible.
These considerations are important for deriving an effective mass interaction
Hamiltonian between the surface electrons and external probes such as magnetic
field or magnetic impurities.
Naturally, this interaction will depend on details of the bulk band structure
of a material.
Below we are going to illustrate this point by comparing the coupling of
surface states and localized magnetic moments in lead and bismuth selenide
compounds.

\begin{figure}[t]
 \begin{center}
  \includegraphics[width = \columnwidth]{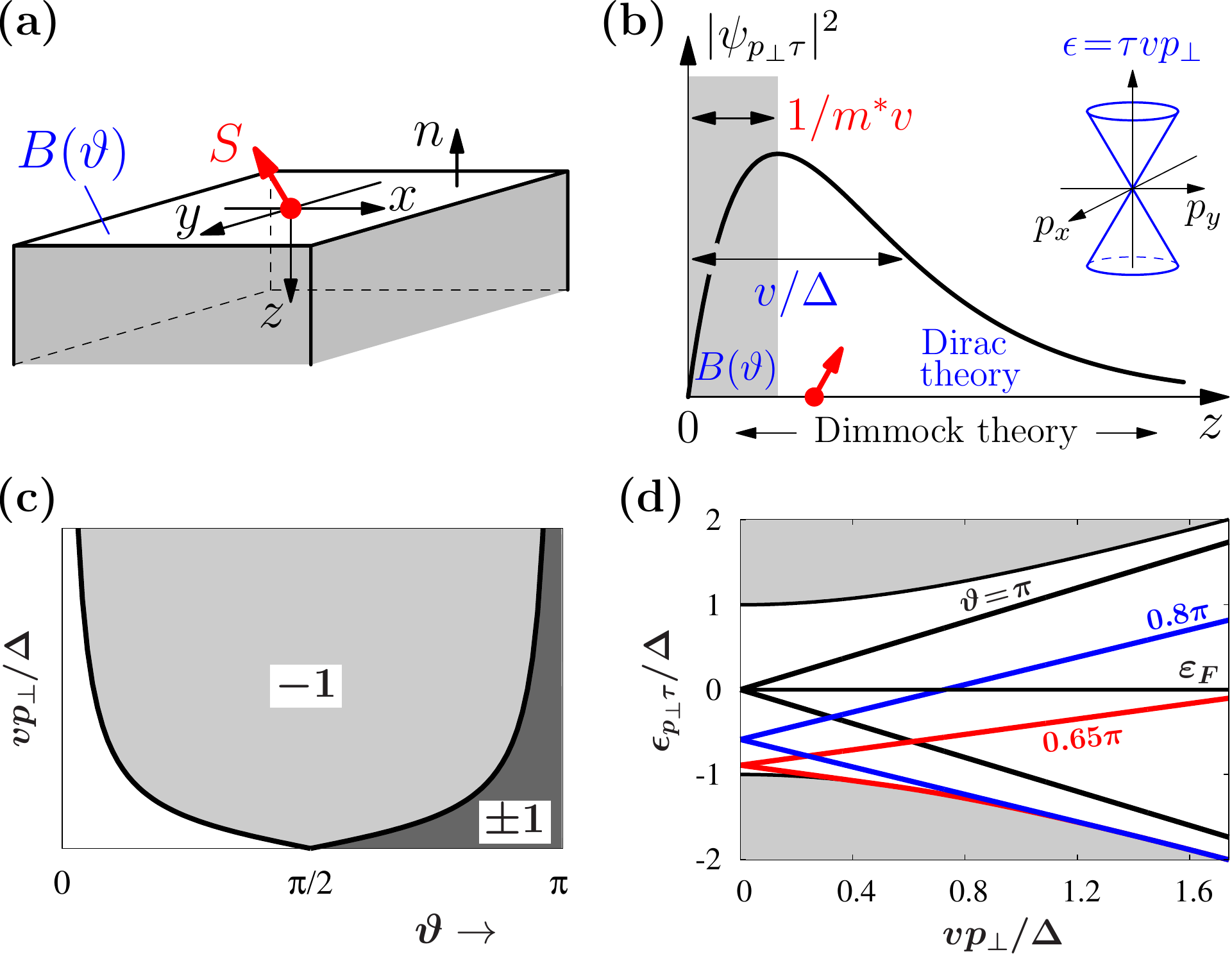}
 \end{center}
 \caption{
  Panel (a) Geometry of the problem.
  The TI occupies half-space $z \geqslant 0$.
  The unit vector ${\bm n}$ is an outer normal to the surface.
  The red arrow represents the impurity spin.
  (b) Schematic $z$-dependence of the surface state wavefunction
  \eqref{eq:dimmock-ss}, characterized by two length scales: small $1 / \lambda
  \sim 1 / m^* v$ (shaded region) and large $1 / q \sim v / \Delta$.
  The Dirac model \eqref{eq:dirac-H} is valid at distances $\geqslant 1 / q$.
  The impurity is located under the surface where the Dirac theory is
  applicable.
  The inset shows the dispersion of surface modes.
  (c) Stability diagram of surface states \eqref{eq:dirac-ss}.
  Thick lines correspond to critical momenta $p_\perp^{cr} = \pm (\Delta / v)
  \, {\rm ctg} \, \vartheta$.
  In the white region no surface states can exist.
  In the (light) dark gray area, there are surface modes with (only one, $\tau
  = -1$) both helicites.
  (d) Dispersion relation \eqref{eq:dirac-ss} for several values of
  $\vartheta$.
  The upper (lower) branches (relative to the point $p_\perp = 0$) correspond
  to $\tau = \mp 1$. The Fermi energy is $\varepsilon_F = 0$.
 }
 \label{fig:fig1}
\end{figure}

\subsection{Quasiparticle states in a half-space}
\label{sec_model_subsec_states}

We are particularly interested in the localized surface states that form as a
result of breaking translational invariance. Consider a TI bounded by the
surface $z = 0$ whose bulk states are described by $H_{\rm D}$, see Fig.
\ref{fig:fig1}(a). It follows that $\bpp$ is conserved and, together with $T^z$,
can be used to classify quasiparticles states. An eigenfunction
$\psi_{\bpp\tau}$ of $T^z$, $T^z\psi_{\bpp\tau}=\tau p_\perp\psi_{\bpp\tau}$,
has the form
\begin{displaymath}
 \psi_{\bpp \tau} ({\bm x}) =
 \begin{pmatrix}
  a (z) U_{\bpp \tau} \\
  b (z) U_{\bpp, -\tau}
 \end{pmatrix}
 \e^{\ii \bpp \cdot {\bm x}_\perp}.
\end{displaymath}
with
\begin{equation}
 U_{\bpp \tau} = \frac{1}{\sqrt{2}}
 \begin{pmatrix}
  1 \\
  -\ii \tau \e^{\ii \phi_\bpp}
 \end{pmatrix},
 \label{eq:U-spinor}
\end{equation}
where $p_\perp = \vert \bpp \vert$ and $\e^{\ii \phi_\bpp}=(p_x + \ii p_y) /
p_\perp$.
The amplitudes $a(z)$ and $b(z)$ are determined by solving the remaining $2
\times 2$ boundary value problem.
In Eq. \eqref{eq:dimmock-H-2x2} one can now replace $T^z$ with $\tau p_\perp$,
hence reducing the number of independent Dirac matrices to two: $\beta$ and
$\alpha^z$.
Their action on the $z$-dependent spinor part of $\psi_{\bpp \tau} ({\bm x})$
is equivalent to the action of $\sigma^z$ and $\sigma^x$ on a two-component
wavefunction $(a^* \, b^*)^\dag$
\begin{align}
 \alpha^z
 \begin{pmatrix}
  a(z) U_{\bpp \tau} \\
  b(z) U_{\bpp, -\tau}
 \end{pmatrix}
 =
 \begin{pmatrix}
  b(z) U_{\bpp \tau} \\
  a(z) U_{\bpp, -\tau}
 \end{pmatrix}
 \to &
 \sigma^x
 \begin{pmatrix}
  a(z) \\
  b(z)
 \end{pmatrix}, \nonumber \\
 \beta
 \begin{pmatrix}
  a(z) U_{\bpp \tau} \\
  b(z) U_{\bpp, -\tau}
 \end{pmatrix}
 =
 \begin{pmatrix}
  a(z) U_{\bpp \tau} \\
  -b(z) U_{\bpp, -\tau}
 \end{pmatrix}
 \to &
 \sigma^z
 \begin{pmatrix}
  a(z) \\
  b(z)
 \end{pmatrix}. \nonumber
\end{align}
which allows us to replace the Hamiltonian \eqref{eq:dimmock-H-2x2} with a $2
\times 2$ operator
\begin{equation}
 H^{(2 \times 2)}_{\rm D} = v \bigl( \sigma^x p_z - \sigma^y \tau p_\perp
 \bigr) + \sigma^z \biggl( \Delta + \frac{{\bm p}^2}{2 m^*} \biggr)
 \label{eq:dimmock-H-2x2_manifest}
\end{equation}
acting on two-component $z$-dependent wavefunctions.
This reduction of dimension (from 4 to 2) is a direct consequence of
conservation of $T^z$.

An important insight can be obtained by studying the simplest case of a hard
boundary at $z = 0$ where the wavefunction vanishes, $\psi \vert_{z = 0} = 0$.
Surface states with energy $\epsilon_{\bpp \tau} = \tau v p_\perp$ exist for an
inverted band structure when $-m^* v^2 / 2 < \Delta < 0$ \cite{kisin-1987-1}.
The surface-state (unnormalized) wavefunction is a coherent superposition of
the conduction and valence bands
\begin{equation}
 \psi_{\bpp \tau} ({\bm x}) \sim
 \begin{pmatrix}
  U_{\bpp \tau} \\
  -\ii U_{\bpp, -\tau}
 \end{pmatrix}
 \bigl( \e^{-q z} - \e^{-\lambda z} \bigr) \e^{\ii \bpp \cdot {\bm x}_\perp},
 \label{eq:dimmock-ss}
\end{equation}
and is characterized by two momentum-dependent inverse length scales:
$\begin{pmatrix} q \\\lambda \end{pmatrix}=m^* v \mp \sqrt{m^* (m^* v^2 + 2
\Delta) + \bpp^2}$.
This surface state is stable only when $q > 0$, i.e. for $p_\perp \leqslant
\sqrt{2 m^* \vert \Delta \vert}$ and merges into the scattering continuum for
larger $p_\perp$.
For more complicated BCs, the problem of determining the surface spectrum from
microscopic considerations is rather cumbersome (see Ref.
\onlinecite{isaev-2011-1} and Appendix \ref{sec_app-a} for details).

It is possible to simplify matters by considering the limit when $m^* v \gg
\vert \Delta \vert / v, \, p_\perp$.
In this case $v q \approx \vert \Delta \vert$ and $\lambda \approx 2 m^* v \gg
q$.
These lengths are illustrated in Fig. \ref{fig:fig1}(b).
One can build a theory \cite{joe-2007-1} valid on the scale $\sim 1 / q$ by
neglecting $1 / \lambda$.
This small-$p_\perp$ perturbative approach is similar to that used in
hydrodynamics of weakly viscous fluids \cite{landau-1963-1}.
In the bulk we can simply omit the ${\bm p}^2 / 2 m^*$ term in Eq.
\eqref{eq:dimmock-H}, so the Hamiltonian takes the Dirac form:
\begin{equation}
 H_0 = v (\balpha \cdot {\bm p}) + \beta \Delta.
 \label{eq:dirac-H}
\end{equation}
Near the surface (at distances $\sim 1 / \lambda$) the situation is more
complicated because the term ${\bm p}^2 / 2 m^* \sim \lambda$ and cannot be
neglected.
Within this layer [shown in gray in Fig. \ref{fig:fig1}(b)] the electronic
wavefunction varies rapidly in accordance with the BCs supplementing
the Dimmock Hamiltonian \eqref{eq:dimmock-H}.
However, this complexity can be absorbed into the BC for the Dirac Hamiltonian
\eqref{eq:dirac-H}.
This BC has to be consistent with the particle conservation, time-reversal and
inversion (parity) symmetries, and can be written as
\cite{volkov-1981-1,satanin-1984-1,isaev-2011-1} $B \psi \vert_{z = 0} = 0$
with
\begin{equation}
 B = 1 + \beta \sin \vartheta + \ii \beta (\balpha \cdot {\bm n}) \cos
 \vartheta,
 \label{eq:dirac-B}
\end{equation}
and ${\bm n}$ -- the outer normal to the surface.
The boundary operator $B$ includes one free parameter $\vartheta$ which
accounts for microscopic properties of a realistic TI surface, and the behavior
of the electronic wavefunction at the length scale $\sim 1 / m^* v$.
An exact connection between $\vartheta$ and the boundary conditions of the
original fully microscopic Hamiltonian is not unique in the effective long
wavelength Dirac model.
However, we show in Appendix \ref{sec_app-a} that to be self-adjoint in a
half-space, the Hamiltonian \eqref{eq:dirac-H} must have a {\it
single-parameter} family of BCs.
Consequently, variation of the parameter $\vartheta$ allows us to consider
entire sets of possible surface properties realized in experiments.
Physically, $\vartheta$ controls the amount of particle-hole (p-h) asymmetry at
the surface: The p-h symmetric case is recovered only when $\vartheta = 0$ or
$\pi$.
The Dirac model \eqref{eq:dirac-H} is clearly less complete than the Dimmock
theory \eqref{eq:dimmock-H}, but it is much easier to work with.

\begin{figure}[t]
 \begin{center}
  \includegraphics[width = \columnwidth]{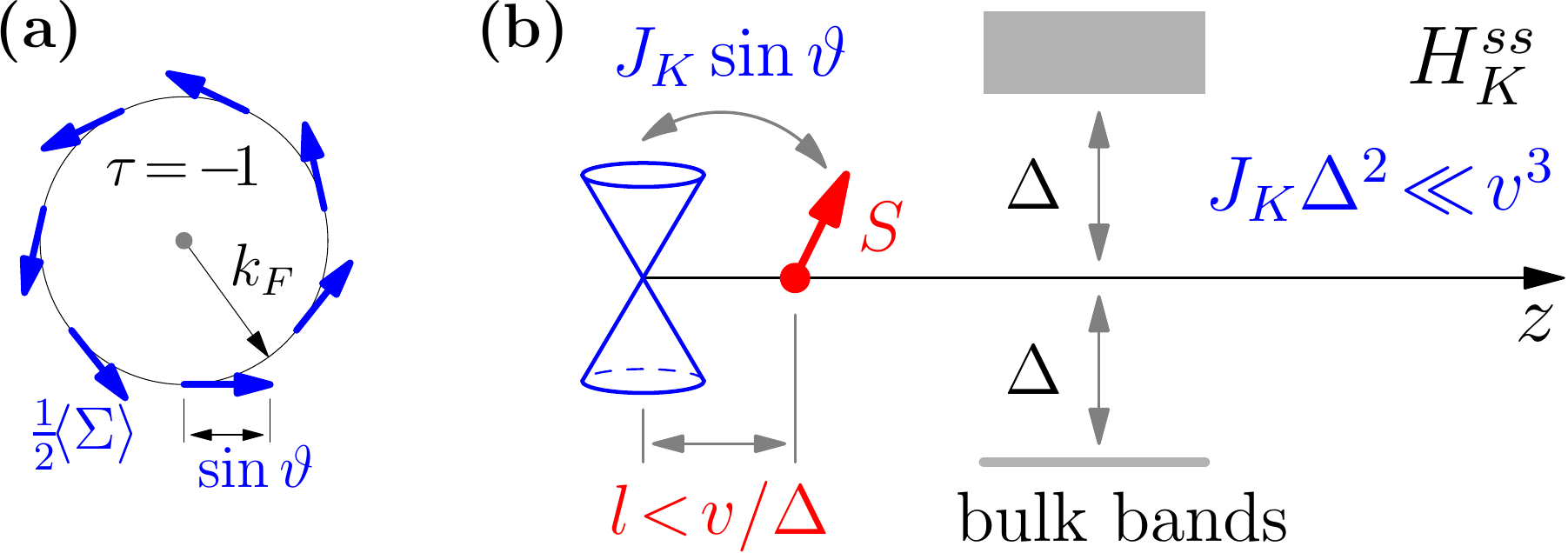}
 \end{center}
 \caption{
  Panel (a) Helical structure of the surface state \eqref{eq:dirac-ss}.
  Blue arrows indicate the expectation value of the electron spin $\frac{1}{2}
  {\bm \Sigma}$ which points perpendicular to the momentum and has a magnitude
  $\sim \sin \vartheta$.
  For $\vartheta = \pi$, surface states carry {\it no spin}.
  (b) Schematic illustration of the Kondo interaction $H_K$ in Eq.
  \eqref{eq:kondo-int-bulk}.
  At weak coupling one can ignore bulk-surface mixing induced by the impurity
  and assume that $H_K \approx H_K^{s s}$.
  The impurity only couples to surface states (blue Dirac cone).
 }
 \label{fig:fig2}
\end{figure}

From now on we will focus on the problem defined by Eqs. \eqref{eq:dirac-H} and
\eqref{eq:dirac-B}.
Since $\vartheta$ can be chosen arbitrarily, we confine our analysis to the
case $\Delta > 0$ (no band inversion) and $\vartheta \in [0, \pi]$.
Results for $\vartheta > \pi$ can be obtained using charge conjugation $\psi
\to \alpha^y \psi^*$.
The energy and wavefunction of surface states are given by
\begin{equation}
 \epsilon_{\bpp\tau}=\tau vp_\perp\cos\vartheta-\Delta\sin\vartheta,
 \label{eq:ss-eps}
\end{equation}
and
\begin{equation}
 \psi_{\bpp\tau} ({\bm x}) = {\cal N}
 \begin{pmatrix}
  (1 - \sin \vartheta) U_{\bpp \tau} \\
  -\ii \cos \vartheta U_{\bpp, -\tau}
 \end{pmatrix} \e^{(\ii \bpp \cdot {\bm x}_\perp - q_{\bpp \tau} z)}.
 \label{eq:dirac-ss}
\end{equation}
Here ${\cal N} = \sqrt{q_{\bpp \tau} / A (1 - \sin \vartheta)}$ and
\begin{equation}
 q_{\bpp\tau}=-\tau p_\perp\sin\vartheta-(\Delta/v)\cos\vartheta
 \label{eq:ss-q}
\end{equation}
is the localization wavevector and $A$ -- the area of the TI surface
[$xy$-plane, see Fig. \ref{fig:fig1}(a)].
The stability region of the state \eqref{eq:dirac-ss} is determined by the
condition $q_{\bpp \tau} \geqslant 0$.
For $\vartheta > \pi / 2$ the $\tau = -1$ state exists for any value of
$p_\perp$, while the state with $\tau = +1$ is stable only for $v p_\perp <
-\Delta \, {\rm ctg} \, \vartheta$.
For $\vartheta < \pi / 2$ the $\tau = +1$ mode is always unstable and the one
with $\tau = -1$ exists for $v p_\perp > \Delta \, {\rm ctg} \, \vartheta$.
These regions are shown in Fig. \ref{fig:fig1}(c).
The surface state enters the single-particle continuum at $p_\perp =
p_\perp^{cr} = \pm (\Delta / v) \, {\rm ctg} \, \vartheta$.
The function $\epsilon_{\bpp\tau}$ is presented in Fig. \ref{fig:fig1}(d) for
several values of $\vartheta$.

Surface states \eqref{eq:dirac-ss} are characterized by a helical spin
distribution, shown in Fig. \ref{fig:fig2}(a), that {\it depends on the BC}, as
one can see by computing an expectation value of the spin $\frac{1}{2} {\bm
\Sigma}$.
This average is $\bigl \langle \frac{1}{2} {\bm \Sigma} \bigr \rangle =
\frac{1}{2} {\cal N}^2 U^\dag_{\bpp \tau} \bigl[ (1 - \sin \vartheta)^2 \,
\bsig + \cos^2 \vartheta \,\, \sigma^z \bsig \sigma^z \bigr] U_{\bpp \tau} =
-q_{\spp \tau} \tau \sin \vartheta \,\, (\sin \phi_\bpp {\bm e}_x - \cos
\phi_\bpp {\bm e}_y) \sim \sin \vartheta$, hence at a p-h symmetric point
$\vartheta = \pi$, surface states \eqref{eq:dirac-ss} [and
\eqref{eq:dimmock-ss}] carry {\it no spin}.
This situation is quite different from the usual case of boundary-independent
surface states \cite{hasan-2010-1,qi-2011-1}.

The single-particle scattering continua are defined by
$\epsilon_{\bpp p_z}=\pm\sqrt{v^2(p_z^2+\bpp^2)+\Delta^2}$ with
$p_z\geqslant0$. Note that $\epsilon_{\bpp p_z}$ is doubly degenerate w.r.t.
$\tau$. The corresponding (unnormalized) wavefunction is
\begin{displaymath}
 \psi_{\bpp p_z \tau} ({\bm x}) \sim \!
 \begin{pmatrix}
  (\Delta+\epsilon_{\bpp p_z})\sin\kappa\,\,U_{\bpp\tau} \\
  -\ii v\bigl[p_z\cos\kappa+\tau p_\perp\sin\kappa\bigr]U_{\bpp,-\tau}
 \end{pmatrix}
 \e^{\ii\bpp\cdot{\bm x}_\perp},
\end{displaymath}
where $\kappa=p_zz+\zeta$ and
\begin{displaymath}
 {\rm tg}\,\zeta=\frac{vp_z\cos\vartheta}{(1+\sin\vartheta)(\Delta+
 \epsilon_{\bpp p_z})-\tau vp_\perp\cos\vartheta}.
\end{displaymath}

Finally, we make two general remarks.
First, for $\vartheta = \pi$ the Dirac surface state \eqref{eq:dirac-ss} has a
structure similar to its Dimmock counterpart \eqref{eq:dimmock-ss}.
An additional negative sign in the lower component of the spinor in Eq.
\eqref{eq:dirac-ss} appears because in the Dimmock theory \eqref{eq:dimmock-H}
we used $\Delta < 0$, while in the Dirac Hamiltonian \eqref{eq:dirac-H}
$\Delta > 0$.
In the latter case the sign of $\Delta$ can be flipped by a unitary rotation
$H_0 \to \Lambda^\dag H_0 \Lambda$ and $B \to \Lambda^\dag B \Lambda$ with
$\Lambda = (\sigma^x \otimes 1)$.
After this transformation the Dirac surface state wavefunction
\eqref{eq:dirac-ss} at $\vartheta = \pi$ becomes identical to Eq.
\eqref{eq:dimmock-ss}.
Hence, conclusions obtained using the Hamiltonian \eqref{eq:dirac-H} should
also be applicable to the Dimmock model.

Second, one has to prove that the Hamiltonian \eqref{eq:dirac-H} is
self-adjoint in the space of wavefunctions satisfying the BC
\eqref{eq:dirac-B}, which is necessary to guarantee that the Dirac model
isphysical and our conclusions can be linked to experimentally observable
quantities.
In Appendix \ref{sec_app-a} we show that this is indeed the case and the BC
\eqref{eq:dirac-B} defines a self-adjoint extension of the Dirac Hamiltonian
\eqref{eq:dirac-H}.

\section{Coupling the topological insulator to surface magnetic impurities}
\label{sec_ss-mag_int}

The Kondo Hamiltonian that describes the interaction of the two electronic
bands with an impurity on the surface at ${\bm x} = {\bm x}_0 = (0, 0, 0)$ has
the form:
\begin{equation}
 H_K = J_K \bS \cdot {\bm s}({\bm x}_0),
 \label{eq:kondo-int-bulk}
\end{equation}
where $\bS$ is the impurity spin and ${\bm s}({\bm x}_0)$ is the electron spin
density at ${\bm x}_0$, the coupling constant $J_K$ is positive (and has units
of energy $\times$ volume).
Based on contributions from different parts of the electron spectrum, the
operator $H_K$ can be decomposed as
\begin{displaymath}
 H_K = {\cal P}_s H_K {\cal P}_s + {\cal P}_b H_K {\cal P}_b + \bigl(
 {\cal P}_s H_K {\cal P}_b + {\rm h.c.} \bigr).
\end{displaymath} 
Here ${\cal P}_s$ (${\cal P}_b$) is the projector on the surface (bulk)
subspace with ${\cal P}_s + {\cal P}_b = 1$.
The first two terms have matrix elements only between surface and bulk
states respectively, the last term describes surface-bulk mixing induced
by the impurity.
Since bulk states are gapped, the pure bulk contribution cannot support Kondo
screening (due to a vanishing density of states at the Fermi surface) and can
be omitted.
For $J_K / l_C^3 \ll \Delta$, with the ``Compton'' length scale $l_C = v /
\Delta$, the off-diagonal surface-bulk mixing term is perturbative, and can be
neglected in a zeroth order approximation [see Fig. \ref{fig:fig2}(b)].
In the following we will focus on the surface term, $H_K^{s s} = {\cal P}_s H_K
{\cal P}_s$.

As already mentioned in Sec. \ref{sec_model_subsec_model}, the relation
between real electron spin and the pseudospin index in the Dirac Hamiltonian is
material-dependent.
We will consider the simplest case of ${\rm PbSe}$-class materials where the
electron spin operator coincides with the pseudospin and has the form
\begin{displaymath}
 {\bm s} ({\bm x}_0) = \frac{1}{2} \cd({\bm x}_0) \, {\bm\Sigma} \,
 c({\bm x}_0) = \frac{1}{2} \cd ({\bm x}_0)
 \begin{pmatrix}
  \bsig & 0 \\
  0 & \bsig
 \end{pmatrix}
 c ({\bm x}_0),
 \label{pb_spin}
\end{displaymath}
where $c ({\bm x})$ is the annihilation operator that corresponds to the
quasiparticle eigenstates \eqref{eq:dirac-ss}, $c ({\bm x}) = \sum_{\bpp \tau}
\psi_{\bpp \tau} ({\bm x}) c_{\bpp \tau} + \textrm{bulk modes}$.

The surface part $H_K^{s s}$ is obtained by computing matrix elements of
${\bm \Sigma}$ between states \eqref{eq:dirac-ss}:
\begin{align}
 \frac{1}{2} & \psi^\dag_{\bppp \tau^\prime} ({\bm x}_0) {\bm \Sigma} \,\,
 \psi_{\bpp \tau} ({\bm x}_0) =
 \frac{\sqrt{q_{p_\perp^\prime \tau^\prime} q_{p_\perp \tau}}}{2 A (1 - \sin
 \vartheta)} \times \nonumber \\
 & \times U^\dag_{\bppp \tau^\prime} \bigl[ (1 - \sin \vartheta)^2 \bsig +
 \cos^2 \vartheta \sigma^z \bsig \sigma^z \bigr] U_{\bpp \tau} = \nonumber \\
 & = \frac{1}{A} \sqrt{q_{p_\perp^\prime \tau^\prime} q_{p_\perp \tau}} \,
 U^\dag_{\bppp \tau^\prime} [-\sin \vartheta \, \bsig^\perp + \sigma^z
 {\bm e}_z] U_{\bpp \tau}, \nonumber
\end{align}
where $\bsig^\perp = \sigma^x {\bm e}_x + \sigma^y {\bm e}_y$ and we used the
identity $U_{\bpp. -\tau} = \sigma^z U_{\bpp \tau}$.
The full effective Hamiltonian is:
\begin{equation}
 H_{\rm ef} = H_0 + H_K^{s s} = \sum_{\bpp \tau} \epsilon_{\bpp \tau}
 \cd_{\bpp \tau} c_{\bpp \tau} + \frac{J_K}{A} \bS \cdot {\bm s}_c,
 \label{eq:PbTe-ss-kim}
\end{equation}
with $\frac{1}{A} {\bm s}_c = {\cal P}_s {\bm s} ({\bm x}_0) {\cal P}_s$:
\begin{displaymath}
 {\bm s}_c = \sum_{\substack{\bppp \tau^\prime \\ \bpp \tau}}
 Q^{p_\perp^\prime \tau^\prime}_{p_\perp \tau}
 \cd_{\bppp \tau^\prime} U^\dag_{\bppp \tau^\prime} [-\sin \vartheta
 \, \bsig^\perp + \sigma^z {\bm e}_z] U_{\bpp \tau} c_{\bpp \tau}
\end{displaymath}
and $Q^{p_\perp^\prime \tau^\prime}_{p_\perp \tau} =
\sqrt{q_{\bppp \tau^\prime} q_{\bpp \tau}}$.
For $\vartheta = \pi$ (hard wall BCs in the Dimmock model) the coupling of
electrons to the impurity spin is purely Ising-type.
Since the impurity spin cannot be dynamically flipped, there is no Kondo effect
in this p-h symmetric case.
This offers the possibility to control the Kondo screening by surface
manipulation via the boundary parameter $\vartheta$.
Even though the bulk Kondo coupling \eqref{eq:kondo-int-bulk} is $SU
(2)$-symmetric, Eq. \eqref{eq:PbTe-ss-kim} describes a Kondo impurity model
with an $XXZ$ exchange anisotropy, which is a direct consequence of the
inversion symmetry breaking at the surface.

Due to factors $Q_{p_\perp \tau}^{p_\perp^\prime \tau^\prime}$ the Hamiltonian
\eqref{eq:PbTe-ss-kim} is equivalent to a Kondo model with spatially non-local
exchange couplings.
This can be seen by rewriting ${\bm s}_c$ in terms of the fermions
$c_{\bpp \alpha} = \sum_\tau (U_{\bpp \tau})_\alpha c_{\bpp \tau}$ with
$\alpha = \ua, \da$:
\begin{align}
 {\bm s}_c = & \sum_{\bppp \bpp} M_{\alpha^\prime \beta^\prime} (\bppp) [-\sin
 \vartheta \, \bsig^\perp + \nonumber \\
 & + \sigma^z {\bm e}_z]_{\beta^\prime \beta} M_{\beta \alpha} (\bpp)
 \cd_{\bppp \alpha^\prime} c_{\bpp \alpha}, \nonumber
\end{align}
where $M_{\alpha \beta} (\bpp) = \sum_\tau \sqrt{q_{p_\perp \tau}}
(U_{\bpp \tau})_\alpha (U^*_{\bpp \tau})_\beta = Q_0 \delta_{\alpha \beta} +
Q_z [\bsig_{\alpha \beta} \times \bpp]_z / p_\perp$ with $Q_0 =
\frac{1}{2} \sum_\tau \sqrt{q_{p_\perp \tau}}$ and $Q_z = \frac{1}{2} \sum_\tau
\tau \sqrt{q_{p_\perp \tau}}$.
When $\vartheta = \pi - \delta \vartheta$ for small $\vert \delta \vartheta
\vert \ll \pi$ and $\frac{v p_\perp}{\Delta} \delta \vartheta \ll 1$, to the
lowest order we have $\frac{Q_z}{Q_0} \approx - \frac{v p_\perp}{2 \Delta}
\delta \vartheta$ and $Q_0
\approx \sqrt{\frac{\Delta}{v}} \, \bigl[ 1 - \bigl( \frac{\delta \vartheta}{2}
\bigr)^2 \bigl( 1 + \frac{v^2 p_\perp^2}{2 \Delta^2} \bigr) \bigr]$, and
\begin{widetext}
 \begin{align}
  {\bm s}_c \approx \sum_{\bppp \bpp} \biggl[ & \frac{\Delta}{v} (-\bsig^\perp
  \delta \vartheta + {\bm e}_z \sigma^z) + \frac{\ii \delta \vartheta}{2}
  {\bm e}_z (\bppp - \bpp) \cdot \bsig + \nonumber \\
  & + \frac{(\delta \vartheta)^2}{2} \bigl \lbrace [(\bppp + \bpp) \times
  {\bm e}_z] + \ii (\bppp - \bpp) \sigma^z \bigr \rbrace -
  \frac{\Delta (\delta \vartheta)^2}{v} {\bm e}_z \sigma^z F_{\bppp \bpp} -
  \frac{\ii v (\delta \vartheta)^2}{4 \Delta} [\bppp \! \times \bpp]
  \biggr]_{\alpha^\prime \alpha} \cd_{\bppp \alpha^\prime}
  c^{\;}_{\bpp \alpha} \,,
  \nonumber
 \end{align}
\end{widetext}
with $2 F_{\bppp \bpp} = 1 + \bigl( \frac{v}{2 \Delta} \bigr)^2
(\bppp + \bpp)^2$.
The first term in this expression will give rise to the usual (local) Kondo
interaction.
The third term describes a purely orbital mechanism to flip the impurity spin
via a non-local $p$-wave coupling with the conduction electrons.
Finally, the longitudinal terms (proportional to ${\bm e}_z$) reflect an
effective Zeeman field originating from electron in-plane motion.

Because $U_{\tau \bpp}$ are eigenstates of $[\bsig\times\bpp]_z$, $H_{\rm ef}$
in Eq. \eqref{eq:PbTe-ss-kim} describes a two-dimensional system
of electrons subjected to a Rashba SOI and interacting with a magnetic
impurity.
From Fig. \ref{fig:fig1}(d) it follows that by tuning $\vartheta$ we can make
one chirality $\tau$ almost completely disappear, which is equivalent to having
a strong SOI dominating single-electron kinetic energy.

The effective model \eqref{eq:PbTe-ss-kim} seems to be incompatible with recent
results \cite{zitko-2010-1,mitchell-2013-1,orignac-2013-1} arguing that there
is always Kondo screening at the surface of a TI.
The root of this discrepancy is the common assumption that TI surface states
can be considered as helical Dirac (or Weyl) fermions.
From Eq. \eqref{eq:ss-eps}, the effective single-particle surface Hamiltonian
has the form $H^{\rm hel}_0 = U_{\bpp \tau} \epsilon_{p_\perp \tau}
U^\dag_{\bpp \tau} = v \cos \vartheta [\bsig \times \bpp]_z - \Delta \sin
\vartheta$, and the usual case encountered in the literature, $H^{\rm hel}_0 =
-v [\bsig \times \bpp]_z$, is recovered when $\vartheta = \pi$.
The above assumption is {\it not universal}: While the free particle dispersion
relation is captured correctly by $H^{\rm hel}_0$, it is non-trivial to couple
these surface electrons to external probes, e.g. impurities or an external
magnetic field.
Interaction terms involving TI surface states have to be derived carefully
taking into account bulk and surface properties, and are material dependent.

Indeed, our results will be completely different for ${\rm Bi_2 Se_3}$.
The tetragonal band structure of this material dictates that an effective mass
expression for the electrons spin is \cite{silvestrov-2012-1}
\begin{displaymath}
 {\bm s}^\prime ({\bm x}_0) = \frac{1}{2} \cd ({\bm x}_0) {\bm \Sigma}^\prime
 c ({\bm x}_0) = \frac{1}{2} \cd ({\bm x}_0)
 \begin{pmatrix}
  \bsig & 0 \\
  0 & \sigma^z \bsig \sigma^z
 \end{pmatrix}
 c ({\bm x}_0).
\end{displaymath}
The cancellation in the spin matrix element which led to the factor $\sin
\vartheta$ in Eq. \eqref{eq:PbTe-ss-kim} does not occur and we
recover an isotropic ($XXX$) Kondo Hamiltonian whose structure is essentially
independent of $\vartheta$:
\begin{displaymath}
 H_K^{ss} = \frac{J_K}{A} \bS \cdot \sum
 Q^{p_\perp^\prime \tau^\prime}_{p_\perp \tau} \, \cd_{\bppp \tau^\prime}
 U^\dag_{\bppp \tau^\prime} \bsig U_{\bpp \tau} c_{\bpp \tau}.
 \label{surface_Hamiltonian_BiSe}
\end{displaymath}
In the {\it particular} case of $\vartheta = \pi$ (when $Q = {\rm const}$) it
is indeed admissible to use the Dirac-Weyl description of surface states with
the Pauli matrices in the effective Hamiltonian being the true electron spin,
as described for example in Ref. \onlinecite{zhang-2009-1}.
However, as indicated above, for ${\rm PbSe}$-class materials this is not the
case.

Another way to experimentally distinguish the above two classes of materials is
by their response to an external homogeneous magnetic field ${\bm h}$ applied
{\it parallel} to the surface.
Without loss of generality we assume that ${\bm h} = h {\bm e}_x$.
The surface electrons couple to this field via a Zeeman term $H_Z =
-\frac{h}{2} \sum_i {\cal P}_s \cd ({\bm x}_i) \Lambda^x c ({\bm x}_i) {\cal
P}_s \bigl \vert_{z = 0}$ with $\Lambda = \Sigma$ or $\Sigma^\prime$.
In ${\rm Bi_2 Se_3}$-like TIs with $\vartheta = \pi$, the full single-particle
Hamiltonian is $H_0 = -\sum_\bpp \cd_{\bpp \alpha} \bigl \{ v [ \bsig \times
\bpp ]_z + (\Delta / v) h \sigma^x \bigr \}_{\alpha \beta} c_{\bpp \beta}$.
Hence, the only effect of $h$ on surface states is to shift the Dirac cone in
the Brillouin zone \cite{fu-2009-1}.
On the contrary, for $\vartheta = \pi$ surface electrons in lead chalcogenides
{\it do not} couple to the transverse field at all, because $s^x_c \equiv 0$.
The Zeeman coupling appears only to the order $(\pi - \vartheta)^2$.
For $\vartheta \neq \pi$, the Dirac-Weyl description of surface states in terms
of $H^{\rm hel}_0$ is meaningless, regardless of the material.

\section{Effective surface Hamiltonian: Orbital nature of screening}
\label{sec_eff-surface-H}

To gain insight into the physical properties of the Kondo impurity model
\eqref{eq:PbTe-ss-kim} we will exploit its axial symmetry which guarantees
conservation of the z-component of the total angular momentum $j_z = l_z +
\frac{1}{2} \Sigma^z$ with $l_z$ being the orbital part.
The fermions $c_{\bpp \tau}$ can be expanded in the angular momentum basis:
\begin{displaymath}
 c_{\bpp \tau} = \sum_m \e^{\ii m \phi_\bpp} c_{p_\perp m \tau}; \quad
 c_{p_\perp m \tau} = \sum_{\phi_\bpp} \e^{-\ii m \phi_\bpp} c_{\bpp \tau},
\end{displaymath}
where the integer $m \in (-\infty, \infty)$.
The sum over $\phi_\bpp$ has to be understood as $\sum_{\phi_\bpp} \to
\int_0^{2 \pi} \frac{d \phi_\bpp} {2 \pi}$.
We also define a sum over the radial momentum $p_\perp$: $\sum_{p_\perp} \to
\frac{A}{2 \pi} \int_0^\infty d p_\perp p_\perp$, so that $\sum_\bpp =
\sum_{p_\perp \phi_\bpp}$.
Moreover, $\delta_{\bppp \bpp} = \delta_{p_\perp^\prime p_\perp}
\delta_{\phi_\bppp \phi_\bpp}$.
Using these relations one can show that $c_{p_\perp m \tau}$ satisfy the
fermionic anticommutation relations $\lbrace \cd_{p_\perp^\prime m^\prime
\tau^\prime}, c_{p_\perp m \tau} \rbrace = \delta_{p_\perp^\prime p_\perp}
\delta_{m^\prime m} \delta_{\tau^\prime \tau}$.

The fermion spin density ${\bm s}_c$ in Eq. \eqref{eq:PbTe-ss-kim} becomes:
\begin{align}
 s_c^+ = & \ii \sin \vartheta \sum_{\substack{p^\prime_\perp \tau^\prime \\
 p_\perp \tau}} Q^{p_\perp^\prime \tau^\prime}_{p_\perp \tau} \tau
 \cd_{p_\perp^\prime 0 \tau^\prime} c_{p_\perp \bar{1} \tau}; \nonumber \\
 s_c^z = & \frac{1}{2} \sum_{\substack{p^\prime_\perp \tau^\prime \\
 p_\perp \tau}} Q^{p_\perp^\prime \tau^\prime}_{p_\perp \tau}
 (\cd_{p_\perp^\prime 0 \tau^\prime} c_{p_\perp 0 \tau} - \tau^\prime \tau
 \cd_{p_\perp^\prime \bar{1} \tau^\prime} c_{p_\perp \bar{1} \tau}). \nonumber
\end{align}
Because only $m = 0$ ($s$-wave) and $m = \bar{1} = -1$ ($p$-wave) angular
harmonics enter these expressions, we can define new fermionic degrees of
freedom \cite{malecki-2007-1}
\begin{equation}
 a_{p_\perp \tau \ua} = c_{p_\perp 0 \tau}, \,\,
 a_{p_\perp \tau \da} = -\ii \tau c_{p_\perp \bar{1} \tau}.
 \label{eq:a-fermions}
\end{equation}
These operators create surface electrons with total angular momentum $j_z = \pm
1 / 2$ (see also Appendix \ref{sec_app-b}).

\begin{figure}[t]
 \begin{center}
  \includegraphics[width = \columnwidth]{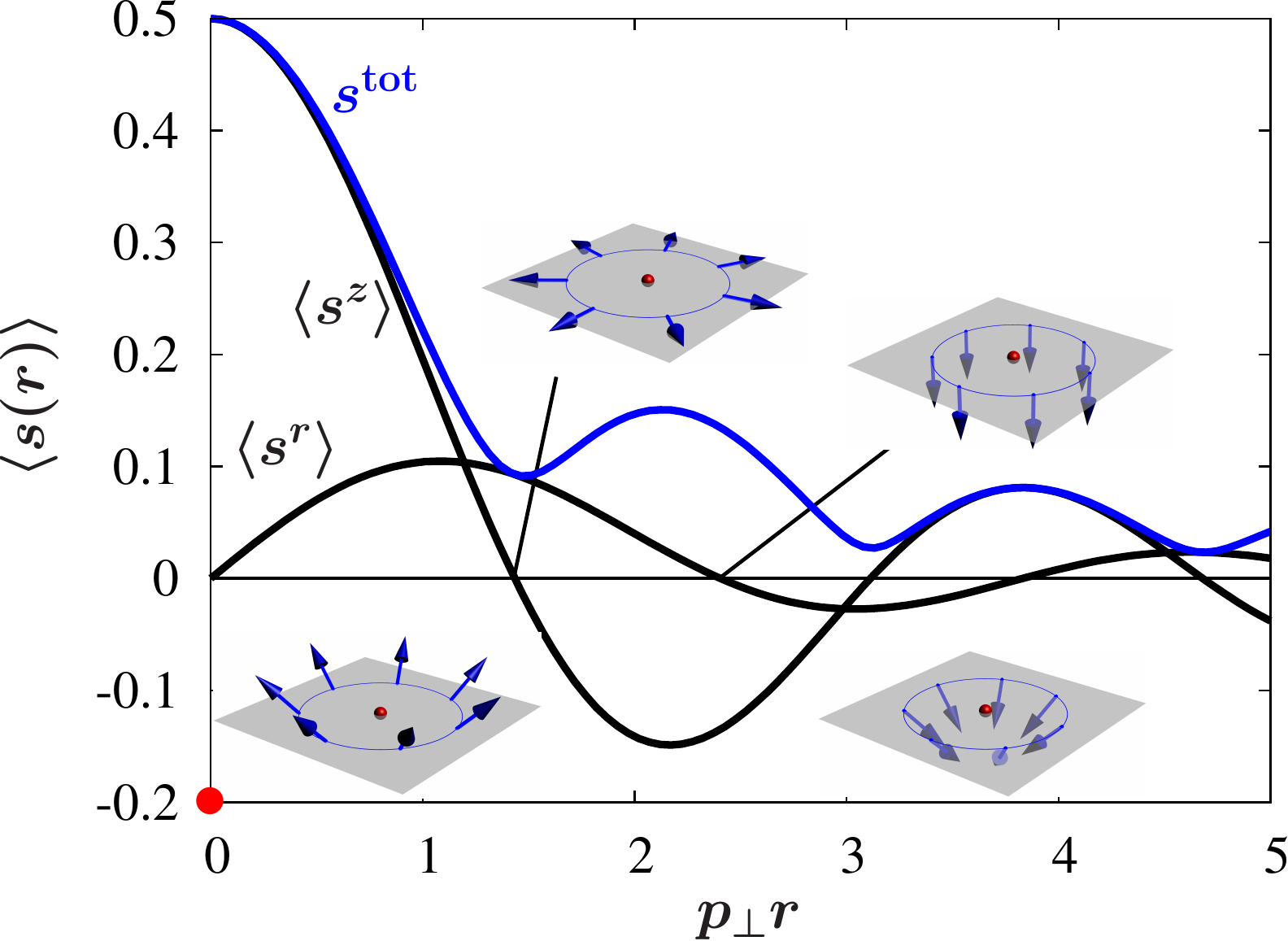}
 \end{center}
 \caption{
  Spin distribution $\langle 1_{\spp \tau \mu} \vert {\bm s} ({\bm x}_\perp)
  \vert 1_{\spp \tau \mu} \rangle = \langle {\bm s} (r) \rangle$, Eq.
  \eqref{eq:mag-vortex} along radial direction with $\tau = \bar{1}$ and
  $\vartheta = 0.9 \pi$.
  $\langle s^r \rangle [\langle s^z \rangle]$ is the radial [z] component.
  The blue line is the ``total'' spin $s^{\rm tot} = \sqrt{\langle s^r
  \rangle^2 + \langle s^z \rangle^2}$.
  Blue arrows show schematic spin distributions at a fixed radius $r$.
  Thick red dots indicate the impurity location at the origin ${\bm x}_0 = (0,
  0, 0)$.
 }
 \label{fig:fig3}
\end{figure}

To give microscopic meaning to the operators \eqref{eq:a-fermions}, it is
instructive to compute the local electron spin density at the surface that
corresponds to a state with one $a$-particle, i.e. an expectation value in the
state $\vert 1_{\spp \tau \mu} \rangle = \ad_{\spp \tau \mu} \vac$ of the
operator
\begin{align}
 {\bm s} ({\bm x}_\perp) = \frac{1}{2}
 \sum_{\substack{\sppp \spp \\ \mu^\prime \mu}}
 \psi^\dag_{\sppp m^\prime \tau} ({\bm x}) {\bm \Sigma} \,
 \psi_{\spp m \tau} ({\bm x}) \biggl \vert_{z = 0} \,\,
 \ad_{\sppp \mu^\prime} a^{\;}_{\spp \mu},
 \nonumber
\end{align}
where $\psi_{\spp m \tau} = \psi_{\spp m (\mu) \tau}$ is the surface state
wavefunction \eqref{eq:dirac-ss} in the angular momentum basis (cf. Appendix
\ref{sec_app-b}), $m^\prime = m (\mu^\prime)$, and $m (\ua) = 0$ and $m (\da) =
-1$.
In the polar coordinates ${\bm x}_\perp = (r \cos \varphi, r \sin \varphi)$, we
have
\begin{align}
 \langle 1_{\spp \tau \mu} & \vert {\bm s} ({\bm x}_\perp) \vert
 1_{\spp \tau \mu} \rangle = \pm \frac{q_{p_\perp \tau}}{A} \times
 \label{eq:mag-vortex} \\
 \times & \biggl \lbrace -\tau \sin \vartheta J_0 (\rho) J_1 (\rho) \,
 {\bm e}_r + \frac{1}{2} [J_0^2 (\rho) - J_1^2 (\rho)] \, {\bm e}_z \biggr
 \rbrace, \nonumber
\end{align}
with ${\bm e}_r = (\cos \varphi, \sin \varphi)$, $\rho = p_\perp r$, and $J_n
(x)$ is the $n$-th Bessel function of the first kind.
The upper (lower) sign corresponds to $\mu = \ua$ ($\da$).
The spin distribution \eqref{eq:mag-vortex} is shown in Fig. \ref{fig:fig3}.
Unlike the plane-wave states $\cd_{\bpp \tau} \vac$, the wavefunctions
$a^\dag_{p_\perp \tau \mu} \vac$ carry no net spin, i.e. $\int d^2 x_\perp
\langle 1_{\spp \tau \mu} \vert {\bm s} ({\bm x}_\perp) \vert 1_{\spp \tau \mu}
\rangle = 0$.

Using operators \eqref{eq:a-fermions}, we can rewrite Eq.
\eqref{eq:PbTe-ss-kim} as
\begin{align}
 H_{\rm ef} = & \sum_{p_\perp \tau} \epsilon_{p_\perp \tau}
 a^\dag_{p_\perp \tau \mu} a_{p_\perp \tau \mu} + \label{eq:PbTe-kim-j} \\
 + \frac{J_K}{2 A} & \sum_{\substack{p^\prime_\perp \tau^\prime \\
 p_\perp \tau}} Q^{p_\perp^\prime \tau^\prime}_{p_\perp \tau} \bS
 \cdot a^\dag_{p_\perp^\prime \tau^\prime \mu^\prime} \bigl( \sin \vartheta \,
 \bsig_{\mu^\prime \mu}^\perp + \sigma^z_{\mu^\prime \mu} {\bm e}_z \bigr)
 a_{p_\perp \tau \mu}. \nonumber
\end{align}
Here we assumed implicit summation over pseudospin indices $\mu$ and
$\mu^\prime = \ua, \da$, omitted all angular harmonics $m \neq 0, \bar{1}$
which do not couple to the impurity, and disregarded the negative sign in the
$XY$-term.
This sign is irrelevant and can be switched by a unitary transformation
$H_{\rm ef} \to U^\dag H_{\rm ef} U$ with $U = 2 S^z$.
Elementary spin-flip scattering processes in Eq. \eqref{eq:PbTe-kim-j}
correspond to dynamical mixing of the spin distributions \eqref{eq:mag-vortex}
and are schematically illustrated in Fig. \ref{fig:fig4}(a) [and should be
contrasted with spin-flip scattering in the usual metal without SOI depicted in
Fig. \ref{fig:fig4}(b)].

The Hamiltonian \eqref{eq:PbTe-kim-j} appears to describe a magnetic impurity
coupled to two conduction bands (channels) labeled by the helicity index
$\tau = \pm 1$.
However, this is not actually the case as can be easily demonstrated by
converting $H_{\rm ef}$ to the energy representation.
We shall consider only energies within the bandgap, $-\Delta \leqslant \epsilon
\leqslant \Delta$ and assume that $\pi / 2 < \vartheta \leqslant \pi$, so $\cos
\vartheta \leqslant 0$.
There is a one-to-one correspondence between $\tau$ and energy (i.e. helicity
of the state and its energy in the upper or lower Dirac cone).
From Eqs. \eqref{eq:ss-eps} and \eqref{eq:ss-q} it follows that $\tau = +1
(-1)$ corresponds to energies $\epsilon < -\Delta \sin \vartheta$ ($\epsilon >
-\Delta \sin \vartheta$).
Since $q_{\bpp \tau}$ and $\epsilon_{\bpp \tau}$ depend only on the product
$\tau p_\perp$,
\begin{equation}
 q (\epsilon) = \frac{\Delta + \epsilon \sin \vartheta}
 {v \vert \cos \vartheta \vert}.
 \label{eq:q-eps}
\end{equation}
Notice that $q (\epsilon) \neq 0$ for all $\epsilon$ within the gap.
Next we derive the density of states (DOS) $g_\tau (\epsilon)$.
For $\tau = +1$ one has $\frac{1}{A} \sum_{p_\perp} =
\int_{-\Delta \sin \vartheta}^{-\Delta} d \epsilon (\epsilon + \Delta \sin
\vartheta) / 2 \pi v^2 \cos^2 \vartheta =
\int^{-\Delta \sin \vartheta}_{-\Delta} d \epsilon \,\, g_+ (\epsilon)$.
Similarly for $\tau = -1$: $\frac{1}{A} \sum_{p_\perp} =
\int_{-\Delta \sin \vartheta}^\Delta d \epsilon \,\, g_- (\epsilon)$ with
$g_- (\epsilon) = (\epsilon + \Delta \sin \vartheta) / 2 \pi v^2 \cos^2
\vartheta$.
Hence for all energies
\begin{equation}
 g (\epsilon) = \frac{\vert \epsilon + \Delta \sin \vartheta \vert}{2 \pi v^2
 \cos^2 \vartheta}.
 \label{eq:ss-dos}
\end{equation}
Finally, we introduce new operators $a_{\epsilon \mu} =
a_{p_\perp (\epsilon) \tau \mu} / \sqrt{g (\epsilon)}$ with anticommutation
relations $\lbrace a^\dag_{\epsilon_\prime \mu^\prime}, a_{\epsilon \mu}
\rbrace = \delta_{\mu^\prime \mu} \delta (\epsilon^\prime - \epsilon)$ which
allow us to reduce $H_{\rm ef}$ to a single-channel form
\begin{align}
 \frac{H_{\rm ef}}{A} = & \int_{-\Delta}^\Delta d \epsilon \,\,
 \epsilon \, a^\dag_{\epsilon \mu} a_{\epsilon \mu} + \frac{1}{2} J_K \bS \cdot
 \int_{-\Delta}^\Delta d \epsilon^\prime \int_{-\Delta}^\Delta d \epsilon
 \times \nonumber \\
 \times & [g (\epsilon^\prime) g (\epsilon) q (\epsilon^\prime)
 q (\epsilon)]^{1 / 2} \, a^\dag_{\epsilon^\prime \mu^\prime} \bigl( \sin
 \vartheta \, \bsig^\perp \! + \sigma^z {\bm e}_z \bigr)_{\mu^\prime \mu}
 a_{\epsilon \mu}. \nonumber
\end{align}
This reduction from a two-channel form \eqref{eq:PbTe-kim-j} occurs because of
the unique correspondence between energy and helicity peculiar to surface
states.

\section{Unconventional Kondo Physics}
\label{sec_results}

The Hamiltonian \eqref{eq:PbTe-kim-j} describes a Kondo impurity model with an
anisotropic ($XXZ$) exchange coupling and a DOS \eqref{eq:ss-dos} that can
vanish at the Fermi level, $\epsilon = 0$, if the BC $\vartheta = \pi$ is
satisfied.
In this limit two effects simultaneously ensure that the Kondo screening does
not occur, and the impurity spin effectively decouples from the surface metal.
First, from numerical renormalization group calculations, for linearly
vanishing DOS and particle-hole symmetry the critical Kondo coupling does not
exist \cite{ingersent-1998-1}.
It is worth noting that this behavior is not captured by the standard mean
field theories \cite{withoff-1990-1}.
Second, in our system the spin-flip scattering is proportional to $\sin
\vartheta$ and therefore disappears at $\vartheta = \pi$.
This effect is already present at the mean field level.
Hence the decoupling of the impurity from the metallic surface states at
$\vartheta = \pi$ is inexorably linked to the anisotropy of the spin scattering
stemming from the bulk band structure.

\begin{figure}[t]
 \begin{center}
  \includegraphics[width = \columnwidth]{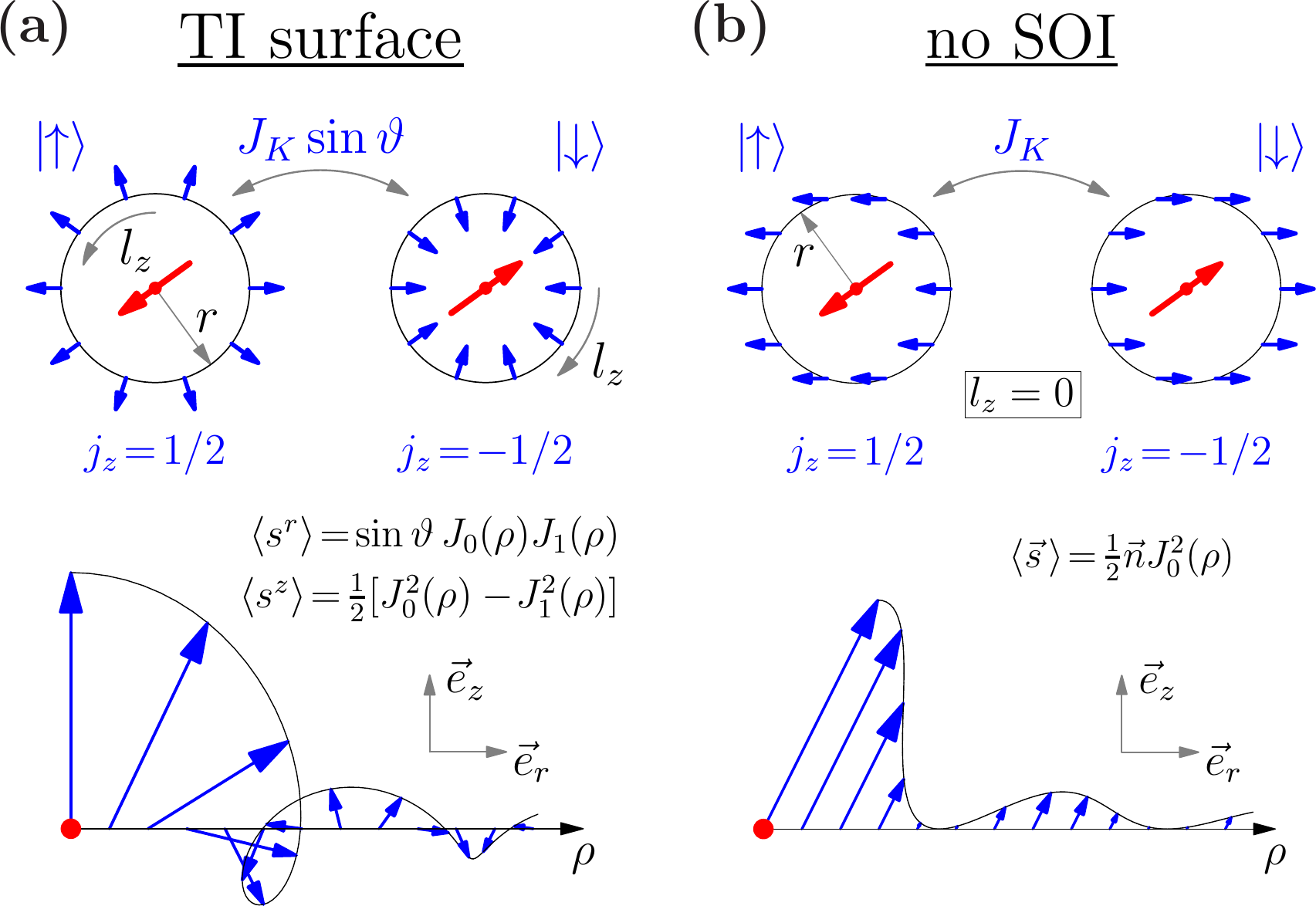}
 \end{center}
 \caption{
  Panel (a) Upper plot: Spin-flip scattering processes leading to the Kondo
  effect.
  The red [blue] arrows denote impurity spin at ${\bm x}_\perp = 0$ [local spin
  \eqref{eq:mag-vortex} in the conduction band at a distance $\vert
  {\bm x}_\perp \vert = r$ from the origin].
  The impurity spin is screened by the orbital degrees of freedom (coupled
  flips of the electron spin and orbital angular momentum $l_z$).
  The lower plot shows a spiral spin structure \eqref{eq:mag-vortex} along a
  radial direction ($\rho = \spp r$) away from the impurity.
  (b) Same as in panel (a) but for a conventional metal without SOI.
  Only conduction electrons in the $s$-wave state couple to the impurity and
  the orbital angular momentum does not participate in the Kondo screening.
  The spin direction (along an arbitrary direction ${\bm n}$) does not depend
  on the radial position.
 }
 \label{fig:fig4}
\end{figure}

For $\frac{\pi}{2} < \vartheta < \pi$ there is a finite DOS \eqref{eq:ss-dos}
at the Fermi surface, and for temperature $T$ below a characteristic Kondo
scale $T_K$, the impurity spin is screened \cite{cox-1999-1,hewson-1997-1}.
This Kondo effect occurs due to the {\it orbital motion} of conduction
electrons [Fig. \ref{fig:fig4}(a)] unlike the conventional case when the
impurity is screened only by itinerant spins [Fig. \ref{fig:fig4}(b)].
More precisely, the impurity spin forms a singlet with the {\it total} angular
momentum ${\bm j}$ of the surface states.
This unconventional mechanism for the Kondo screening originates from the
strong SOI that couples spin and orbital momentum of electrons in TIs.

In the following we would like to address the physical manifestations of this
unconventional Kondo effect.
We first demonstrate the appearance of a transverse spin linear response to a
longitudinal external magnetic field.
We next consider the effect of temperature and study the dependence of the
Kondo temperature on the electronic surface properties parameterized by
$\vartheta$.
Although we focus on the model \eqref{eq:PbTe-kim-j} obtained in the context of
TIs, results of the present section are applicable to Kondo physics in any
two-dimensional metal with SOI.

\subsection{Transverse local magnetic response}

\begin{figure}[t]
 \begin{center}
  \includegraphics[width = \columnwidth]{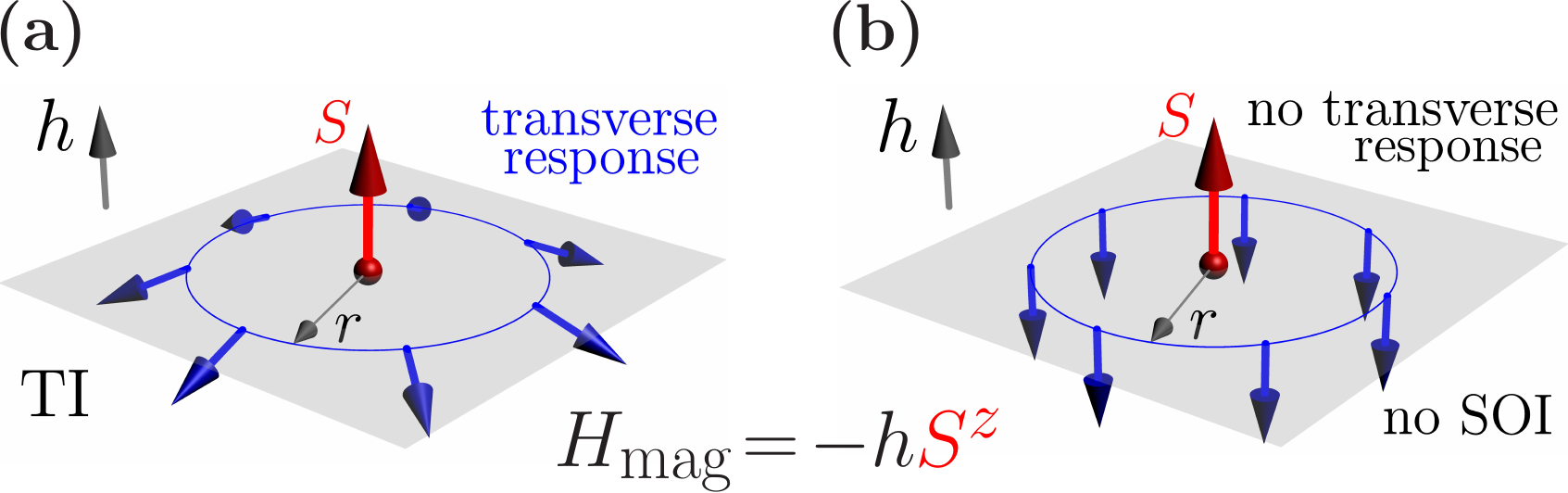}
 \end{center}
 \caption{
  Panel (a) Schematic plot of the transverse magnetic response on a TI surface.
  A magnetic field $h$ applied normal to the surface causes a radial electron
  spin polarization.
  (b) Same as panel (a), but for the Kondo effect in a usual two-dimensional
  metal without SOI.
  There is only longitudinal magnetic response.
 }
 \label{fig:fig5}
\end{figure}

The simplest manifestation of the spin-orbital nature of the Kondo effect on a
TI surface can be found in the zero temperature ($T = 0$) linear response to a
weak magnetic field ${\bm h}$ acting on the impurity.
Assuming that ${\bm h} = h {\bm e}_z$ points perpendicular to the surface, the
field correction to the model \eqref{eq:PbTe-kim-j} is
\begin{displaymath}
 H_{\rm mag} = -h S^z.
 \label{zeeman_term}
\end{displaymath}
According to Eq. \eqref{eq:mag-vortex}, surface states with $\mu = \ua$ and
$\da$ correspond to different (opposite) {\it radial} spin distributions.
In the Kondo singlet state both configurations are equally probable and the
total spin in the $xy$-plane vanishes.
However, in an applied magnetic field the impurity spin is weakly polarized
creating a population imbalance of electrons with different $\mu$'s.
This imbalance results in a {\it transverse} local (i.e. at a fixed distance
from the impurity) spin polarization in the {\it conduction band}, see Fig.
\ref{fig:fig5}(a).

To calculate the field-induced transverse magnetization we use the standard
variational approach \cite{yosida-1966-1,isaev-2011-1} for the Kondo problem,
and assume that $\vartheta < \pi$ so that all $\tau = +1$ states are filled and
the Fermi level lies in the $\tau = -1$ cone in Fig. \ref{fig:fig1}(d).
At weak coupling $\frac{J_K \Delta^2}{v^3} \ll 1$ one needs to keep only $\tau
= -1$ terms in Eq. \eqref{eq:PbTe-kim-j}, hence in the rest of this Subsection
we will omit $\tau$ in the subscripts.
The variational wavefunction has the form \cite{ishii-1967-1}:
\begin{equation}
 \vert \psi_0 \rangle = \sum_{\spp \geqslant k_F} [{\cal A}_\spp
 \chi^s_{\mu \alpha} + {\cal B}_\spp \chi^t_{\mu \alpha}] \ad_{\spp \mu} \FS
 \otimes \vert \alpha \rangle,
 \label{eq:var-state}
\end{equation}
where $k_F = \frac{\Delta \sin \vartheta}{v \vert \cos \vartheta \vert}$ is the
Fermi momentum, and $\FS$ and $\vert \alpha \rangle$ are the Fermi sea and
impurity spin states ($\alpha = \ua, \da$) respectively.
There is an implicit summation over spin indices.
The two terms in \eqref{eq:var-state} correspond to singlet ($\chi^s$) and
triplet ($\chi^t$) components with $\chi^{s, t}_{\mu \alpha} =
\frac{1}{\sqrt{2}} (\delta_{\mu \ua}^{\alpha \da} \mp
\delta_{\mu \da}^{\alpha \ua})$.
The latter satisfy the relations $\chi^s_{\mu \alpha} \chi^s_{\mu \alpha} =
\chi^t_{\mu \alpha} \chi^t_{\mu \alpha} = 1$, $\chi^s_{\mu \alpha}
\chi^t_{\mu \alpha} = 0$, and $S^z_{\alpha \beta} \chi^s_{\mu \beta} =
\frac{1}{2} \chi^t_{\mu \alpha}$.
The state \eqref{eq:var-state} is normalized according to $\langle \psi_0
\vert \psi_0 \rangle = \sum_{\spp \geqslant k_F} \bigl( \vert {\cal A}_\spp
\vert^2 + \vert {\cal B}_\spp \vert^2 \bigr) = 1$.

The amplitudes ${\cal A}_\spp$ and ${\cal B}_\spp$ are variational parameters
determined by minimizing the functional ${\cal F} = \langle \psi_0 \vert
H_{\rm ef} + H_{\rm mag} \vert \psi_0 \rangle - (E_{\rm FS} - \lambda) \langle
\psi_0 \vert \psi_0 \rangle$, where $\lambda$ is the Lagrange multiplier that
plays the role of an energy shift due to the Kondo screening.
To the first order in $h$, a straightforward calculation yields
\begin{displaymath}
 \begin{pmatrix}
  {\cal A}_\spp \\ {\cal B}_\spp
 \end{pmatrix}
 = \frac{J_K (1 + 2 \sin \vartheta)}{4 A}
 \frac{c \sqrt{q_\spp}}{(\epsilon_\spp + \lambda)^2}
 \begin{pmatrix}
  \epsilon_\spp + \lambda \\ -h / 2
 \end{pmatrix},
\end{displaymath}
with $\spp \geqslant k_F$ and $c = \sum_{\spp \geqslant k_F} \sqrt{q_\spp}
A_\spp$.
The eigenvalue $\lambda$ is determined from the non-linear equation
\begin{equation}
 1 = \frac{J_K (1 + 2 \sin \vartheta)}{4 A} \sum_{\spp \geqslant k_F}
 \frac{q_\spp}{\epsilon_\spp + \lambda}.
 \label{eq:yosida-eq}
\end{equation}
At weak coupling the sum can be computed as $\frac{1}{A} \sum_\spp \ldots =
\int_0^\Delta d \epsilon \,\, \frac{g (\epsilon) q (\epsilon)}{\epsilon +
\lambda} \approx q (0) g (0) \ln \frac{\Delta}{\lambda}$ [$q (\epsilon)$ is
given in Eq. \eqref{eq:q-eps}] which means that $\ln \frac{\lambda}{\Delta}
\approx -8 \pi v^3 \vert \cos^3 \vartheta \vert / \Delta^2 J_K \sin \vartheta
(1 + 2 \sin \vartheta)$.
Then, the normalization constant $c$ is given by $c^2 = \frac{A \lambda}{g(0)
q(0)} \bigl[ \frac{4}{J_K (1 + 2 \sin \vartheta)} \bigr]^2$.

If, as is commonly done, one identifies the energy shift $\lambda$ with the
Kondo temperature $T_K$, we find that, as $\vartheta \rightarrow \pi$, $T_K$
vanishes exponentially as $T_K \sim \Delta \exp \bigl[-8 \pi v^3 / \Delta^2 J_K
(\pi - \vartheta) \bigr]$.
Note, however, that in this approach for a finite DOS at the Fermi level the
variational energy shift does not vanish when spin-flip processes are
suppressed.
Consequently, in next subsection we take this effect into account and define
the Kondo temperature using the slave-boson method. 

The field-induced transverse spin distribution in the ground state $\vert
\psi_0 \rangle$ is straightforwardly obtained using Eq.
\eqref{eq:polar-s-matrix} and the discussion in Sec. \ref{sec_eff-surface-H}:
\begin{align}
 \langle \psi_0 \vert s^r ({\bm x}_\perp) \vert \psi_0 \rangle = & \frac{\sin
 \vartheta}{A} \sigma^x_{m n} \times \nonumber \\
 \times \sum_{\sppp \spp \geqslant k_F} & \sqrt{q_\sppp q_\spp}
 {\cal A}_\sppp {\cal B}_\spp J_n (\sppp r) J_m (\spp r),
 \nonumber
\end{align}
with $n, m = 0$ and $1$.
With the aid of the above expressions for $\lambda$, $c$, ${\cal A}_\spp$ and
${\cal B}_\spp$, we finally arrive at
\begin{displaymath}
 \langle \psi_0 \vert s^r ({\bm x}_\perp) \vert \psi_0 \rangle =
 -\frac{4 h \sin \vartheta}{(1 + 2 \sin \vartheta) J_K} J_0 (k_F r) J_1
 (k_F r),
 \label{transverse_polarization}
\end{displaymath}
where we also employed a weak-coupling approximation for the energy integrals
$\int_0^\Delta d \epsilon f (\epsilon) / (\epsilon + \lambda)^n \approx f (0)
\int_0^\Delta d \epsilon / (\epsilon + \lambda)^n$ ($f$ is a smooth function
and $n \geqslant 0$ is an integer).
Due to the structure of the variational state \eqref{eq:var-state} the spin
distribution is identical up to a prefactor to Eq. \eqref{eq:mag-vortex} with
$\spp = k_F$ [see also Fig. \ref{fig:fig3}].

The transverse magnetic response, i.e. nonzero $\langle s^r \rangle \sim h$,
can be viewed as a variation of the Edelstein effect
\cite{aronov-1989-1,edelstein-1990-1}: an applied magnetic field creates an
imbalance of different orbital angular momentum states that couple to the
impurity, which in turn induces a radial spin polarization.
This phenomenon exists only due to the SOI and is absent in metals without SOI
[see Fig. \ref{fig:fig5}(b)].
Therefore, by studying the spatial structure of the Kondo resonance, for
example by spin-polarized STM, one can differentiate between topologically
non-trivial and trivial states of matter.
Field-induced radial spin spirals similar to Figs. \ref{fig:fig4}(a) and
\ref{fig:fig5}(a) were reported in Ref. \onlinecite{chirla-2013-1} in
connection to Kondo screening of magnetic impurities on gold surfaces with a
weak Rashba SOI $\alpha_R$.
In that work, the transverse susceptibility $\kappa_\perp = \langle s^r
\rangle / h \sim \alpha_R$.
Our results deal with an opposite limit of strong SOI and hence $\kappa_\perp$
depends only on $J_K$ and the boundary parameter $\vartheta$.

In the absence of an external field $h$, the ground state wavefunction
\eqref{eq:var-state} is an $SU (2)$-singlet, despite the $XXZ$ anisotropy of
the Kondo model \eqref{eq:PbTe-kim-j}.
This is an example of the general irrelevance of exchange anisotropies for the
Kondo physics \cite{cox-1999-1}.
However, in our case this emergent $SU (2)$ symmetry is quite non-trivial
because the impurity spin forms a singlet with the {\it total angular momentum}
of the surface electrons [see Fig. \ref{fig:fig4}(a)].
Coupling to the orbital motion ensures that this singlet-formation is the
physical mechanism responsible for the Kondo resonance even when electron spins
are quenched by the strong SOI.

\subsection{Slave-boson mean-field approach}

In the previous Subsection we assumed that for any $\vartheta < \pi$ the
impurity is screened by surface electrons with only one helicity $\tau = -1$.
Here we verify this conjecture by studying the model in Eq.
\eqref{eq:PbTe-kim-j} within the slave boson mean-field approach
\cite{hewson-1997-1,orignac-2013-1}.
This analysis also provides an extension of our previous results to finite
temperature.

\begin{figure}[t]
 \begin{center}
  \includegraphics[width = 0.9 \columnwidth]{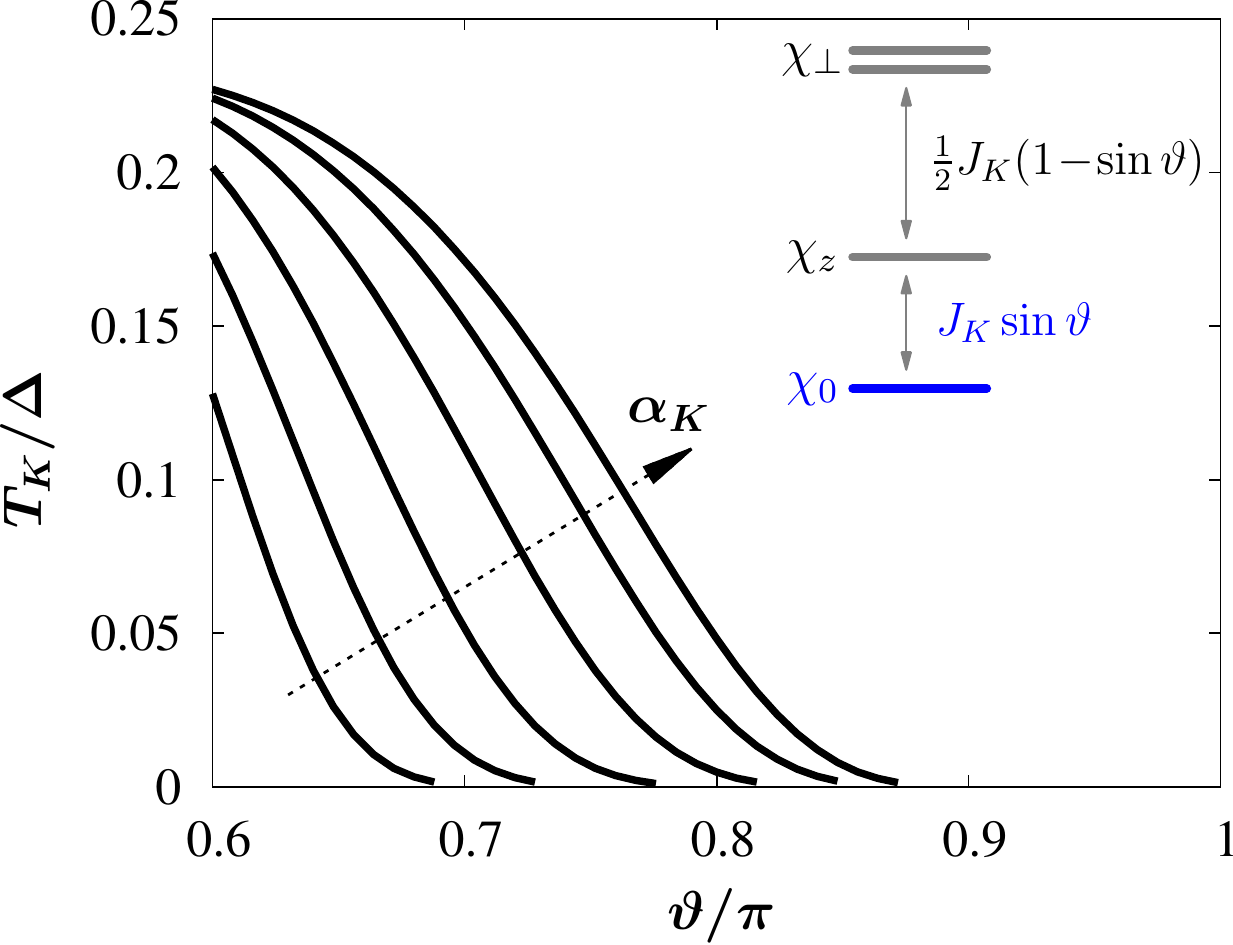}
 \end{center}
 \caption{
  Kondo temperature, computed from the Nagaoka-Suhl equation
  \eqref{eq:nagaoka-eq}, as a function of the BC angle $\vartheta$ [see Eq
  \eqref{eq:dirac-B}].
  The arrow shows increasing values of the dimensionless Kondo coupling
  $\alpha_K = J_K \Delta^2 / 2 \pi v^3 = 0.05$, $0.1$, $0.2$, $0.4$, $0.7$,
  $1.0$.
  Inset: Energies involved in the Hamiltonian \eqref{eq:chi-bosons}.
  The blue color indicates the condensed slave boson $\chi_0$.
 }
 \label{fig:fig6}
\end{figure}

First we introduce a pseudofermion representation of the local spin $\bS =
\frac{1}{2} \fd_\mu \bsig_{\mu \nu} f_\nu$ with the constraint $\sum_\mu
\fd_\mu f_\mu = 1$.
In this language the interaction term in $H_{\rm ef}$ can be written in a
compact form
\begin{align}
 H_{\rm ef} = & -J_K \sin \vartheta \hat{\chi}^\dag_0 \hat{\chi_0} + J_K
 \frac{1 - \sin \vartheta}{2} \hat{\bm \chi}^\dag_\perp \cdot
 \hat{\bm \chi}_\perp - \nonumber \\
 & - J_K \frac{1 - 2 \sin \vartheta}{4}
 \sum_{\substack{\sppp \tau^\prime \\ \spp \tau}}
 Q^{\sppp \tau^\prime}_{\spp \tau} \ad_{\sppp \tau^\prime \mu}
 a_{\spp \tau \mu}.
 \label{eq:chi-bosons}
\end{align}
The slave bosons are defined as \cite{kusminskiy-2008-1}: $\hat{\chi}_l =
\frac{1}{\sqrt{2 A}} \sum_{\spp \tau} \sqrt{q_{\spp \tau}} \fd_\mu
\sigma^l_{\mu \nu} a^{\;}_{\spp \tau \nu}$ with $l = 0, 1,2, 3$ and
$\sigma^0_{\mu \nu} = \delta_{\mu \nu}$ [see Fig. \ref{fig:fig6}].
Notice, that the zero energy in Eq. \eqref{eq:chi-bosons} is chosen such that
it eliminates $\hat{\chi}_z$, which is necessary since energies of the states
with condensed $\chi_0$ and $\chi_z$ bosons (see below) are only different when
spin-flip scattering is present.
This procedure adds a potential scattering term that preserves the impurity
spin and is therefore irrelevant for the Kondo physics [cf Ref.
\onlinecite{orignac-2013-1}].

The mean-field appoximation amounts to treating the pseudofermion constraint on
the average via a chemical potential $E_f$, and assuming that the ground
state corresponds to condensation of the $\chi_0$ boson, i.e. $\langle
\hat{\bm \chi}_\perp \rangle = 0$ but $\langle {\hat \chi}_0 \rangle = \chi_0
\neq 0$.
The mean-field Hamiltonian,
\begin{align}
 H_{\rm MF} = & \sum_{\spp \tau \mu} \epsilon_{\spp \tau} \ad_{\spp \tau \mu}
 a_{\spp \tau \mu} - E_f \sum_\mu \fd_\mu f_\mu - \nonumber \\
 & - \frac{J_K \sin \vartheta}{2 A} \sum_{\spp \tau \mu} \sqrt{q_{\spp \tau}}
 \bigl( \chi_0 \ad_{\spp \tau \mu} f_\mu + {\rm h.c.} \bigr), \nonumber
\end{align}
can be diagonalized using the equations of motion method for retarted Green
functions \cite{zubarev-1960-1,nagaoka-1965-1} which, for fermions, are defined
as $\langle \langle A; B \rangle \rangle =  -\ii \theta(t - t^\prime) \langle
\lbrace A (t), B(t^\prime) \rbrace \rangle$ [$\theta(x)$ is the Heaviside
step function].
We will need three types of Green functions: $\langle \langle
a^{\;}_{\spp \tau \mu}; \ad_{\sppp \tau^\prime \mu} \rangle \rangle$, $\langle
\langle f^{\;}_\mu; \fd_\mu \rangle \rangle$, and $\langle \langle f^{\;}_\mu;
\ad_{\spp \tau \mu} \rangle \rangle$.
A direct calculation yields:
\begin{align}
 \langle a^{\;}_{\spp \tau \mu}; & \ad_{\sppp \tau^\prime \mu} \rangle_\omega =
 \frac{\delta_{\sppp \spp}^{\tau^\prime \tau}}{\omega - \epsilon_{\spp \tau}} -
 \nonumber \\
 & - \frac{J_K \sin \vartheta}{\sqrt{2 A}} \frac{\chi_0 \sqrt{q_{\spp \tau}}}
 {\omega - \epsilon_{\spp \tau}} \langle f^{\;}_\mu;
 \ad_{\sppp \tau^\prime \mu} \rangle_\omega \nonumber \\
 \langle f^{\;}_\mu; & \ad_{\spp \tau \mu} \rangle_\omega = -\frac{J_K \sin
 \vartheta \chi_0^* \sqrt{q_{\spp \tau}}}{\sqrt{2 A} (\omega -
 \epsilon_{\spp \tau}) \bigl[ \omega + E_f - \Omega(\omega) \bigr]}, \nonumber
\end{align}
where we introduced the Fourier transform
$\langle \langle A; B \rangle \rangle = \frac{1}{2 \pi} \int_{-\infty}^\infty
d \omega \, \e^{-\ii \omega (t - t^\prime)} \langle A; B \rangle_\omega$, and
the impurity self-energy
\begin{displaymath}
 \Omega(\omega) = \frac{\vert J_K \sin \vartheta \chi_0 \vert^2}{2 A}
 \sum_{\spp \tau} \frac{q_{\spp \tau}}{\omega - \epsilon_{\spp \tau}}.
\end{displaymath}
The mixed Green function $\langle f^{\;}_\mu; \ad_{\spp \tau \mu}
\rangle_\omega$ allows us to construct the self-consistency equation for
$\chi_0$:
\begin{align}
 \chi_0^* = & \frac{1}{\sqrt{2 A}} \sum_{\spp \tau \mu} \sqrt{q_{\spp \tau}}
 \langle \ad_{\spp \tau \mu} f^{\;}_\mu \rangle = \nonumber \\
 & = \frac{1}{\sqrt{2 A}} \sum_{\spp \tau \mu} \sqrt{q_{\spp \tau}} \int d
 \omega \, {\cal A}_{\spp \tau \mu} (\omega), \nonumber
\end{align}
with the spectral function
\begin{displaymath}
 {\cal A}_{\spp \tau \mu}(\omega) = \frac{\ii}{2 \pi} \frac{\langle f^{\;}_\mu;
 \ad_{\spp \tau \mu} \rangle_{\omega + \ii 0} - \langle f^{\;}_\mu;
 \ad_{\spp \tau \mu} \rangle_{\omega - \ii 0}}{\e^{\omega / T} + 1}.
\end{displaymath}

At the Kondo temperature $T_K$, defined by $E_f(T_K) = \Omega(T_K) = 0$, the
above self-consistency condition reduces to the Nagaoka-Suhl equation
\begin{equation}
 1 = -\frac{J_K \sin \vartheta}{A} \,\, {\rm P.V.} \sum_{\spp \tau}
 \frac{q_{\spp \tau}}{\epsilon_{\spp \tau} \bigl(
 \e^{\epsilon_{\spp \tau} / T_K} + 1 \bigr)},
 \label{eq:nagaoka-eq}
\end{equation}
where ${\rm P.V.}$ indicates the Cauchy principal value.
This expression generalizes the Yosida equation \eqref{eq:yosida-eq} for the
case where both helicities $\tau$ are allowed to participate in the Kondo
screening.

For $\vartheta$ sufficiently distinct from $\pi$ and weak coupling we only need
to consider conduction band states near the Fermi energy.
All of them have the same helicity due to the one-to-one correspondence between
$\tau = \pm 1$ and energy leading to the helicity-independent DOS
\eqref{eq:ss-dos}.
This fact justifies our assumptions made in the previous Subsection.

Beyond weak coupling, the sum in \eqref{eq:nagaoka-eq} can be computed
numerically using Eqs. \eqref{eq:q-eps} and \eqref{eq:ss-dos} for $q
(\epsilon)$ and $g (\epsilon)$.
The dependence of the Kondo temperature on $\vartheta$ is shown in Fig.
\ref{fig:fig6}.
Asymptotically, for $\vartheta \to \pi$, the Kondo temperature $T_K\sim \Delta
\exp \bigl[-2 \pi v^3 / \Delta^2 J_K (\pi - \vartheta)^2 \bigr]$ is
exponentially suppressed, albeit its functional behavior is different from that
obtained using the variational approach.
Since at $\vartheta = \pi$, the $XY$ term in \eqref{eq:PbTe-kim-j} vanishes,
there is no critical Kondo coupling that would yield a finite $T_K$ at this
point [cf. Ref. \onlinecite{withoff-1990-1}].

\section{Discussion}
\label{sec_discussion}

In the present work we advocated the use of magnetic probes to test and tune
the unconventional phenomena at topological insulator surfaces.
We showed that physical characteristics and quantum numbers of the surface
states are quite sensitive to surface properties encoded in boundary conditions
for electron wavefunctions, as well as the structure of bulk Bloch bands.
Moreover, we demonstrated how the combination of spin-orbit interaction and
non-trivial boundary conditions leads to an unconventional Kondo screening of
dilute magnetic impurities on the surface of a 3D topological insulator.

We considered a localized spin $S = \frac{1}{2}$ (magnetic) impurity atom
deposited on the $(1 1 1)$ surface of a ${\rm Pb Te}$-class narrow-band
semiconductor, and derived a low-energy effective theory that governs the
coupling of this local spin to surface electrons taking into account the full
3D structure of surface-state wavefunctions.
The resulting Kondo impurity model is spatially non-local and anisotropic
[$XXZ$-like, see Eq. \eqref{eq:PbTe-kim-j}].
Interestingly, both of these features are controlled by parameters defined by
the boundary conditions, in our case $\vartheta$, that determine the magnitude
of the particle-hole asymmetry at the surface [see Fig. 1(d)].
Specifically, at the particle-hole symmetric point $\vartheta = \pi$ the $XY$
component of the Kondo exchange interaction vanishes, signalling an {\it
instability of the Kondo screened ground state} (for any amount of surface
gating) due to the lack of spin-flip processes.

When the particle-hole symmetry {\it is broken by the boundary conditions}, we
find that the impurity spin is fully screened by the surface electrons, in
agreement with earlier works \cite{zitko-2010-1,feng-2010-1,orignac-2013-1,
mitchell-2013-1,tran-2010-1,zitko-2011-1}.
However, unlike the conventional Kondo effect \cite{hewson-1997-1}, here the
local spin forms a singlet with the {\it total angular momentum} of itinerant
electrons (as opposed to only their spin) and is screened mainly by the {\it
orbital} electronic degrees of freedom.
This effect originates in the strong spin-orbit interaction that underpins the
helical structure of the surface states, and manifests itself in a {\it
transverse} spin response: A weak, normal to the surface, magnetic field
induces an {\it in-plane} electron spin polarization [see Fig.
\ref{fig:fig5}(a)] which locally resembles a $q = 1$ magnetic vortex (see Ref.
\onlinecite{nomura-2010-1}) with itinerant spins aligning along the radial
direction.

The sensitivity of the Kondo screening to specific surface properties shows
that it is {\it impossible} to provide a universal theory of topological
insulator surface states based solely on topological arguments
\cite{qi-2011-1,hasan-2010-1} without involving knowledge of the boundary
conditions for the Bloch states (see Sec. \ref{sec_model} and Ref.
\onlinecite{isaev-2011-1}), and, as elaborated in the present work, the
specific bulk band structure.
Most importantly, the latter defines the {\it set of relevant effective
operators} that parameterize the surface theory.
Indeed, in Sec. \ref{sec_ss-mag_int} we demonstrated that for a
${\rm Bi_2 Se_3}$-like tetragonal material the form of the surface Kondo
interaction is completely different (isotropic, $XXX$-like) than in cubic
${\rm Pb Te}$-like systems (anisotropic, $XXZ$-like).

This {\it physical} non-universality of topological surface states can be
exploited in experimental studies of topological insulators, for instance to
control the surface spin polarization with external electric and magnetic
fields.
Although we focused on magnetic impurities, our analysis can be generalized to
any magnetic interaction, e.g. the Zeeman coupling of surface electrons to
external fields.
For ${\rm Bi_2 Se_3}$-like materials, the only effect of an in-plane magnetic
field is to shift (neglecting the Fermi surface warping) the Dirac cone in the
Brillouin zone \cite{fu-2009-1}.
However, in ${\rm Pb Te}$-like crystals with a particle-hole symmetric boundary
($\vartheta = \pi$) such field {\it does not} couple to surface states at all.
In general, this coupling can be tuned by surface manipulation.
The above result shows a convenient way of discriminating between different
types of topological insulators by using interactions of surface states with
external magnetic probes.

The transverse spin structures in Fig. \ref{fig:fig5}(a) can be observed in
scanning tunneling microscopy measurements of the local spin-polarized density
of states around the impurity, or nuclear magnetic resonance experiments.
This predicted effect is not peculiar to topological insulators and should in
fact exist in any strong spin-orbit coupled metallic host.
A similar idea of probing the local spin polarization around magnetic
impurities in a metal without spin-orbit interaction, i.e. the analysis of the
Kondo screening cloud, was discussed before \cite{affleck-2010-1}.
Unlike our analysis, in that work the magnetic field induced only a
longitudinal (and no transverse) spin polarization [Fig. \ref{fig:fig5}(b)].

Finally, we comment on the role of impurity charge fluctuations in multiband
Dirac-like materials with strong spin-orbit coupling.
The standard Kondo impurity Hamiltonian is typically derived from the Anderson
impurity model via a Schrieffer-Wolff (SW) transformation assuming that charge
fluctuations at the impurity get suppressed \cite{hewson-1997-1}.
In the absence of spin-orbit interaction, the virtual transitions included in
the SW transformation preserve the electron spin quantum number.
The effective Kondo exchange then depends on momentum only via energy, and near
the Fermi level can be approximated by a constant value.
This situation may change in a spin-orbit coupled system when the electron
transitions between local and itinerant states are accompanied by a spin-flip.
For a ${\rm Pb Te}$-like host these processes can be captured with a modified
3D Anderson impurity model
\begin{displaymath}
 H_{\rm AIM} \! = \! H_0 + \frac{1}{\sqrt{N}} \sum_\bp \bigl[
 V^c_{\alpha \beta} (\bp) \cd_{\bp \alpha} d^{\;}_\beta + V^v_{\alpha \beta}
 (\bp) h^\dag_{\bp \alpha} d^{\;}_\beta + {\rm h.c.} \bigr],
\end{displaymath}
written in terms of the fermion operators $d^\dag_\alpha$,
$c^\dag_{\bp \alpha}$ and $h^\dag_{\bp \alpha}$ that create electrons in the
impurity orbital, conduction and valence band, respectively.
$H_0 = H_{\rm D} + H_d$, with $H_{\rm D}$ from Eq. \eqref{eq:dimmock-H} and
$H_d$ representing the self-energy of the localized electrons.
The matrix amplitudes $V^c_{\alpha \beta}$ and $V^v_{\alpha \beta}$ describe
hybridizations of the local impurity level with electrons in the conduction and
valence bands.

A phenomenological form of these amplitudes can be obtained from general
symmetry considerations.
We require that $H_{\rm AIM}$ has the same symmetries as the non-interacting
Dimmock model $H_{\rm D}$, in particular, time-reversal invariance and symmetry
w.r.t. spatial inversion ${\cal P}$.
The former demands that $V^c$ and $V^v$ contain spin (via the Pauli matrices)
and momentum in even power combinations, e.g. $p^2$ or $\bsig \cdot \bp$.
The inversion symmetry dictates which of these terms actually occur in each
hybridization amplitude.
Under ${\cal P}$, local fermions are invariant $d_\alpha \stackrel{\cal P}{\to}
d_{\alpha}$, while conduction and valence band electrons transform as
\cite{landau-vol-4} $c_{\bp \alpha} \stackrel{\cal P}{\to} c_{-\bp \alpha}$ and
$h_{\bp \alpha} \stackrel{\cal P}{\to} -h_{-\bp \alpha}$.
This means that $V^c$ ($V^v$) is an even (odd) function of $\bp$: $V^c (\bp) =
V_{c 0} + V_{c 1} p^2 + \cdots$ and $V^v (\bp) = V_{v 1} (\bsig \cdot \bp) +
\cdots$.
To lowest order in momentum, $V^c$ can be taken $\bp$-independent:
$V^c_{\alpha \beta} (\bp) \approx V_{c 0} \delta_{\alpha \beta}$.
On the other hand, $V^v_{\alpha \beta} (\bp) \approx V_{v 1}
(\bsig_{\alpha \beta} \cdot \bp)$ has a $p$-wave structure and is spatially
non-local.
This result differs from the calculations of Refs.
\onlinecite{lu-2013-1,kuzmenko-2014-1} which used constant values for both
amplitudes $V^c$ and $V^v$.

Applying the SW transformation to $H_{\rm AIM}$ yields a modified effective
Kondo model: apart from the local exchange coupling $J_K$, there are
essentially non-local corrections that include $p$-wave couplings between
conduction electrons and the local impurity spin.
We considered the simplest version in this paper and leave the more complex
situation for a future investigation.

\section{Acknowledgements}

L.I. was supported by the NSF (PIF-1211914 and PFC-1125844), AFOSR, AFOSR-MURI,
NIST and ARO individual investigator awards, and also in part by ICAM.
I.V. acknowledges support from NSF Grants DMR-1105339 and DMR-1410741.

\appendix
\section{Self-adjoint extensions of the Dirac and Dimmock Hamiltonians in the
half-space}
\label{sec_app-a}

Given a linear bounded operator ${\bm O}$, its adjoint ${\bm O}^\dag$ is
defined as $\langle \psi \vert {\bm O}^\dag \phi \rangle = \langle {\bm O} \psi
\vert \phi \rangle$ for all vectors $\vert \psi \rangle$ and $\vert \phi
\rangle$ in the Hilbert space ${\cal H}$.
Moreover, ${\bm O}$ is symmetric (or Hermitian) if $\langle \psi \vert {\bm O}
\phi \rangle = \langle {\bm O} \psi \vert \phi \rangle$ or ${\bm O}^\dag =
{\bm O}$ for all vectors $\vert \psi \rangle$ and $\vert \phi \rangle$.
The set of all vectors $\vert \phi \rangle$ for which ${\bm O} \vert \phi
\rangle$ is defined is called the domain of the operator ${\bm O}$.
For a bounded symmetric operator ${\bm O}$ its domain covers the entire space:
${\cal D} ({\bm O}) = {\cal D} ({\bm O}^\dag) = {\cal H}$.

On the other hand, if a linear operator ${\bm H}$ is unbounded its domain does
not necessarily coincide with that of its adjoint.
One can make these two domains coincide by defining them appropriately.
If ${\cal D} ({\bm H}^\dag)$ contains ${\cal D} ({\bm H})$, and in ${\cal D}
({\bm H})$ the two operators are the same, then we say that ${\bm H}^\dag$ is
an extension of ${\bm H}$.
A symmetric operator ${\bm H}$ with a dense domain is {\it self-adjoint}
whenever ${\cal D} ({\bm H}) = {\cal D} ({\bm H}^\dag)$ \cite{ahari-2015-1}.

In this section we prove that the Dirac Hamiltonian \eqref{eq:dirac-H}
in the half-space $z \geqslant 0$ is self-adjoint in the domain of
wavefunctions satisfying the BC \eqref{eq:dirac-B} (the following analysis can
also be seen as another derivation of this BC). We also determine self-adjoint
extensions (SAEs) of the Dimmock Hamiltonian \eqref{eq:dimmock-H} in the
half-space. This constitutes a crucial step to discussing and analyzing surface
or interface phenomena that is physically observable. 

The general theory of self-adjoint extensions can be found, for instance, in
Ref. \onlinecite{gitman-2012-1}.
Its practical application to an operator $H$, however, is rather
straightforward \cite{ahari-2015-1,bonneau-2001-1,araujo-2004-1} and was made
systematic by von Neumann's method of deficiency indices.
First, one constructs deficiency subspaces of the adjoint operator, i.e.
determines eigenfunctions $\psi_\pm$ of $H^\dag$ corresponding to eigenvalues
$\pm \ii \eta$ with arbitrary $\eta > 0$.
Dimensions of these subspaces, the deficiency indices $n_\pm$, give the number
of parameters needed to construct families of possible SAEs: if $n_+ = n_- = n
= 0$ the operator is already self-adjoint, otherwise ($n > 0$) its extensions
need to be built.
When $n_+ \neq n_-$ the operator cannot be made self-adjoint.

Provided $n_+ = n_-$, the next step is to demand that the positive and
negative deficiency subspaces be unitarily related by a $n \times n$ matrix
$U$. This matrix is arbitrary and therefore the number of possible SAEs is
$n^2$. Finally, we require that the combination $\psi_+ + U \psi_-$ belong to
the domain of the original operator $H$. This yields BC for the wavefunctions
that define the domain in which $H$ is self-adjoint. The arbitrary unitary
matrix $U$ represents {\it all} possible BCs compatible with $H$ being
self-adjoint. One can then consider SAEs that are constrained by additional
symmetry conditions, such as time-reversal invariance or parity.

\subsection{Dirac Hamiltonian in the half-space $z \geqslant 0$}

For purely imaginary eigenvalues $\pm \ii \eta$ of the Hamiltonian
\eqref{eq:dirac-H} it follows that $p_z = \pm \ii \sqrt{(\eta^2 +
\Delta^2) / v^2 + p_\perp^2} = \pm \ii \kappa$.
To find the corresponding eigenfunctions $\psi_\pm$, it is convenient to reduce
Eq. \eqref{eq:dirac-H} to a $2 \times 2$ form similar to Eq.
\eqref{eq:dimmock-H-2x2_manifest}:
\begin{displaymath}
 H^{(2 \times 2)} = v \bigl( \sigma^x p_z - \sigma^y \tau p_\perp \bigr) +
 \sigma^z \Delta,
\end{displaymath}
and make a unitary transformation generated by
\begin{displaymath}
 \zeta = \frac{1}{\sqrt{2}}
 \begin{pmatrix}
  1 & 1 \\
  \ii & -\ii
 \end{pmatrix}.
\end{displaymath}
so that $\zeta^\dag \bsig \zeta = (\sigma^y, \sigma^z, \sigma^x)$. In this
representation:
\begin{displaymath}
 \psi_\pm =
 \begin{pmatrix}
  \pm \ii \eta - v \tau p_\perp \\
  \Delta - v \kappa
 \end{pmatrix}
 \e^{-\kappa z},
\end{displaymath}
where we dropped the unimportant, for the analysis below, dependence on
${\bm x}_\perp$ as well as the normalization constant (which is the same for
$\psi_+$ and $\psi_-$). Clearly, this solution exists for any sign of
$\pm \ii \eta$, hence the deficiency indices are $n_+ = n_- = 1$. By the von
Neumann theorem, the Hamiltonian \eqref{eq:dirac-H} has a single-parameter, $n
= 1$, family of SAEs.
The unitary matrix, connecting $\psi_+$ and $\psi_-$ is just $\e^{\ii \lambda}$
with an arbitrary $\lambda$.

Possible SAEs are found in the form of BC for a general wavefunction
$\varphi = (\varphi_1^*, \varphi_2^*)^\dag$ from the domain of $H$. The
condition that $H$ is self-adjoint if
\begin{displaymath}
 \langle \psi \vert H \varphi \rangle - \langle H^\dag \psi \vert \varphi
 \rangle = - \ii \psi^\dag \alpha^z \varphi \bigl \vert_{z = 0} = 0
\end{displaymath}
[$\psi \in {\cal D} (H^\dag)$ and we are interested in functions such that
${\cal D} (H) = {\cal D} (H^\dag)$]. Substituting
$\psi = \psi_+ + \e^{\ii \lambda} \psi_-$, we obtain (note that $\alpha^z$ is
equivalent to $\sigma^y$):
\begin{displaymath}
 \frac{\varphi_1}{\varphi_2} \biggl \vert_{z = 0} = -\frac{v \tau p_\perp \cos
 \lambda / 2 - \eta \sin \lambda / 2}{(\Delta - v \kappa) \cos \lambda / 2} =
 \rho,
\end{displaymath}
where $\rho$ is an arbitrary real constant.

As a final step, we would like to recast this BC in a form
$B^\prime \varphi \vert_{z = 0} = 0$ ($\det B^\prime = 0$). The matrix
$B^\prime$ can be written as
\begin{align}
 B^\prime = & b_1
 \begin{pmatrix}
  b_2 & b_2 \rho \\
  1 & \rho
 \end{pmatrix}
 = \frac{b_1}{2} \bigl[ (b_2 + \rho) + (b_2 - \rho) \sigma^z + \nonumber \\
 & \qquad \qquad + (1 + b_2 \rho) \sigma^x - \ii (1 - b_2 \rho) \sigma^y
 \bigr], \nonumber
\end{align}
with arbitrary real $b_2$ (notice that $b_1$ is irrelevant).
This BC preserves time-reversal and parity
invariance of the Dirac Hamiltonian. The BC of Eq. \eqref{eq:dirac-B} is
recovered after inverting the $\zeta$-transformation, i.e. replacing $\bsig$
with $(\beta, \alpha^z, -\ii \beta \alpha^z)$, and taking
$b_1 = 2 \rho / (1 + \rho^2)$ and $b_2 = 1 / \rho$. Then
$\sin \vartheta = 2 \rho / (1 + \rho^2)$ and
$\cos \vartheta = (1 - \rho^2) / (1 + \rho^2)$.

\subsection{Dimmock Hamiltonian in the half-space $z \geqslant 0$}

Similarly to the previous subsection, for the Dimmock model
\eqref{eq:dimmock-H}, we have:
\begin{displaymath}
 \frac{p_z^2}{2 (m^* v)^2} = -\biggl[ 1 + \frac{\Delta + p^2_\perp / 2 m^*}
 {m^* v^2} \biggr] \pm \sqrt{1 + \frac{2 \Delta - \eta^2 / m^* v^2}{m^* v^2}}.
\end{displaymath}
It is straightforward to check that for any values of the parameters $\Delta$
and $p_\perp$, there are two normalizable solutions that decay with $z \to
\infty$. Therefore, the deficiency indices are $n_+ = n_- = 2$, and the
self-adjoint extension of Eq. \eqref{eq:dimmock-H} is realized by a
four-parametric family of BCs.

\section{Surface states in the total angular momentum basis}
\label{sec_app-b}

The eigenvalue problem defined by the Dirac Hamiltonian
\eqref{eq:dirac-H} and its BC \eqref{eq:dirac-B} has an axial symmetry
around the $z$-axis which leads to conservation of the $z$-component of the
total angular momentum $j_z = l_z + \frac{1}{2} \Sigma^z$ ($l_z$ is the orbital
angular momentum).
Here we will employ this symmetry to construct surface states with a definite
value of $j_z$, and derive their spin structure \eqref{eq:mag-vortex} and
coupling to the impurity [see Eq. \eqref{eq:PbTe-kim-j}].

We will work in cylindrical coordinates $(r, \varphi, z)$ with $0 \leqslant r <
\infty$ and $0 \leqslant \varphi < 2 \pi$, related to the Cartesian basis in
Fig. \ref{fig:fig1}(a) via $x = r \cos \varphi$ and $y = r \sin \varphi$.
The vector product $[\bsig \times \bpp]_z$ entering the tensor spin operator
$T_z$ [see Eq. \eqref{eq:tensor-spin-T}] has the form
\begin{displaymath}
 [\bsig  \times \bpp]_z =
 \begin{pmatrix}
  0 & \e^{-\ii \varphi} \bigl( \frac{\partial}{\partial r} + \frac{l_z}{r}
  \bigr) \\
  \e^{\ii \varphi} \bigl( -\frac{\partial}{\partial r} + \frac{l_z}{r}
  \bigr) & 0
 \end{pmatrix},
\end{displaymath}
with $l_z = -\ii \frac{\partial}{\partial \varphi}$.
The eigenstates of this operator are
\begin{displaymath}
 U_{p_\perp m \tau} = \frac{1}{\sqrt{2 A}}
 \begin{pmatrix}
  J_m (p_\perp r) \e^{\ii m \varphi} \\
  \tau J_{m + 1} (p_\perp r) \e^{\ii (m + 1) \varphi}
 \end{pmatrix}.
\end{displaymath}
Here $p_\perp = \vert \bpp \vert$ and $J_m(x)$ is the Bessel function of the
first kind, of order $m$.
This wavefunction is analogous to Eq. \eqref{eq:U-spinor} with $\bpp$
replaced by a pair $(p_\perp, m)$. It is normalized to the total surface area
$A$:
\begin{align}
 \int & d^2 x_\perp U^\dag_{p^\prime_\perp m^\prime \tau^\prime}
 U_{p_\perp m \tau} = \frac{2 \pi \delta_{m^\prime m}}{2 A}
 \frac{\delta (p^\prime_\perp \! - p_\perp)}{\sqrt{p_\perp^\prime p_\perp}}
 (1 + \tau^\prime \tau) \to \nonumber \\
 & \to \frac{2 \pi \delta_{m^\prime m}}{2 A} \biggl( \frac{A}{2 \pi}
 \delta_{p^\prime_\perp p_\perp} \biggr) (2 \delta_{\tau^\prime \tau}) =
 \delta_{p^\prime_\perp p_\perp} \delta_{m^\prime m} \delta_{\tau^\prime \tau},
 \nonumber
\end{align}
where we used the relation between discrete and continuous (Dirac)
$\delta$-functions, $\delta_{p_\perp^\prime p_\perp} \to \frac{2 \pi}{A}
(p_\perp^\prime p_\perp)^{-1 / 2} \delta (p_\perp - p_\perp^\prime)$ and
$\delta_{\phi_\bppp \phi_\bpp} \to 2 \pi \delta (\phi_\bppp - \phi_\bpp)$ that
follow from the vector relation $\delta_{\bppp \bpp} =
\delta_{p_\perp^\prime p_\perp} \delta_{\phi_\bppp \phi_\bpp} \to
\frac{(2 \pi)^2}{A} \delta (\bpp - \bppp) = \frac{(2 \pi)^2}{A} \delta
(\phi_\bppp - \phi_\bpp) (p^\prime_\perp p_\perp)^{-1 / 2} \delta
(p^\prime_\perp - p_\perp)$ (see also the discussion at the beginning of Sec.
\ref{sec_eff-surface-H}).
There is also a completeness relation $\sum_{p_\perp m \tau}
[U^*_{p_\perp m \tau} (r^\prime, \varphi^\prime)]_\alpha [U_{p_\perp m \tau}
(r, \varphi)]_\beta = \delta_{\alpha \beta} \delta ({\bm x}_\perp^\prime -
{\bm x}_\perp)$.
Using well-known properties of the Bessel functions \cite{batygin-1970-1}, we
can relate $U_{\spp m \tau}$ and plane-wave spinors of Eq. \eqref{eq:U-spinor}:
\begin{equation}
 \frac{\e^{\ii \bpp \cdot {\bm x}_\perp}}{\sqrt{A}} U_{\bpp \tau} =
 \sum_{m = -\infty}^\infty \ii^m \e^{-\ii m \phi_\bpp} U_{\spp m \tau} (r,
 \varphi).
 \label{eq:U-spinor-expansion}
\end{equation}

The surface-state wavefunction \eqref{eq:dirac-ss} can be written as
$\psi_{\bpp \tau} ({\bm x}_\perp) = \sum_m \ii^m \e^{-\ii m \phi_\bpp}
\psi_{\spp m \tau}$ with
\begin{displaymath}
 \psi_{\spp m \tau} = \sqrt{\frac{q_{\spp \tau}}{1 - \sin \vartheta}}
 \begin{pmatrix}
  (1 - \sin \vartheta) U_{p_\perp m \tau} (r, \varphi) \\
  -\ii \cos \vartheta \, U_{p_\perp m, -\tau} (r, \varphi)
 \end{pmatrix}
 \e^{-q_{p_\perp \tau} z}.
\end{displaymath}
It is easy to show that $j_z \psi_{\spp m \tau} = \bigl( m + \frac{1}{2} \bigr)
\psi_{\spp m \tau}$.
Furthermore, the operator ${\bm s}_c$ from Eq. \eqref{eq:PbTe-ss-kim} becomes
\begin{align}
 {\bm s}_c = & \frac{1}{2} \sum_{\substack{p^\prime_\perp m^\prime
 \tau^\prime \\ p_\perp m \tau}} Q_{p_\perp \tau}^{p_\perp^\prime \tau^\prime}
 C^\dag_{p_\perp^\prime m^\prime \tau^\prime} \bigl \lbrace -\sin \vartheta
 \bigl[ ({\bm e}_x + \ii {\bm e}_y) \tau^\prime
 \delta_{m 0}^{m^\prime \bar{1}} + \nonumber \\
 & + ({\bm e}_x - \ii {\bm e}_y) \tau \delta_{m \bar{1}}^{m^\prime 0} \bigr] +
 {\bm e}_z \bigl( \delta_{m 0}^{m^\prime 0} - \tau^\prime \tau
 \delta_{m \bar{1}}^{m^\prime \bar{1}} \bigr) \bigr \rbrace
 C_{p_\perp m \tau} = \nonumber \\
 = & \frac{1}{2} \sum_{\substack{p^\prime_\perp \tau^\prime \\ p_\perp \tau}}
 Q_{p_\perp \tau}^{p_\perp^\prime \tau^\prime} (-\sin \vartheta \,
 \bsig^\perp + {\bm e}_z \sigma^z)_{\mu^\prime \mu}
 \ad_{p_\perp^\prime \tau^\prime \mu^\prime} a_{p_\perp \tau \mu}, \nonumber
\end{align}
where we used the fact that $J_m (0) = \delta_{m 0}$ and $C_{\spp m \tau} =
\ii^m c_{\spp m \tau}$.
The operators $a_{p_\perp m \tau}$ are defined via $a_{p_\perp \tau \ua} =
C_{p_\perp 0 \tau}$ and $a_{p_\perp \tau \da} = \tau C_{p_\perp \bar{1} \tau}$.
The latter expression differs from the analogous definition in Eq.
\eqref{eq:a-fermions} by a pure phase $-\ii$ which can be tracked to the above
relation between fermion operators $C_{\spp m \tau}$ and $c_{\spp m \tau}$, as
well as the factor $\ii^m$ in the expansion \eqref{eq:U-spinor-expansion}.
When plugged into the Kondo Hamiltonian \eqref{eq:PbTe-ss-kim}, the above
expression will yield the model \eqref{eq:PbTe-kim-j}.

To compute spatial spin distributions we will need the matrix element
\begin{align}
 {\bm s}^{p_\perp^\prime m \tau}_{p_\perp m \tau} & ({\bm x}_\perp, z = 0) =
 \frac{1}{2} \psi^\dag_{p_\perp^\prime m \tau} {\bm \Sigma}
 \psi_{p_\perp m \tau} \biggl \vert_{z = 0} = \nonumber \\
 = & \frac{\sqrt{q_{p_\perp^\prime \tau} q_{p_\perp \tau}}}{A} \biggl \lbrace
 -\tau \sin \vartheta \, G_m (\rho, \rho^\prime) {\bm e}_r + \zeta_m (\rho,
 \rho^\prime) {\bm e}_z + \nonumber \\
 & \qquad \qquad \qquad + \ii \tau \sin \vartheta \, F_m (\rho, \rho^\prime)
 {\bm e}_\varphi \biggr \rbrace, \nonumber
\end{align}
where $\rho = p_\perp r$, $\rho^\prime = p_\perp r^\prime$, ${\bm e}_r =
{\bm e}_x \cos \varphi + {\bm e}_y \sin \varphi$, ${\bm e}_\varphi =
-{\bm e}_x \sin \varphi + {\bm e}_y \cos \varphi$, and
\begin{displaymath}
 \begin{pmatrix}
  G_m \\ F_m \\ \zeta_m
 \end{pmatrix}
 = \frac{1}{2}
 \begin{pmatrix}
  J_m (\rho^\prime) J_{m + 1} (\rho) \pm J_m (\rho) J_{m + 1} (\rho^\prime) \\
  J_m (\rho^\prime) J_m (\rho) - J_{m + 1} (\rho^\prime) J_{m + 1} (\rho) \\
 \end{pmatrix}.
\end{displaymath}
Importantly, $G_m$ and $\zeta_m$ are symmetric w.r.t. interchange of their
arguments [$G_m (\rho, \rho^\prime) = G_m (\rho^\prime, \rho)$ and $\zeta_m
(\rho, \rho^\prime) = \zeta_m (\rho^\prime, \rho)$], while $F_m$ is
antisymmetric [$F_m (\rho, \rho^\prime) = -F_m (\rho^\prime, \rho)$].
We will only consider the case $m = 0$ and $\bar{1}$.
By virtue of the relation $J_{-1} (\rho) = -J_1 (\rho)$, $G_0 =
-G_{\bar{1}}$, $\zeta_0 = -\zeta_{\bar{1}}$ and $F_0 = F_{\bar{1}}$, and we get
\begin{align}
 {\bm s}^{p_\perp^\prime m \tau}_{p_\perp m \tau} = &
 \frac{\sqrt{q_{p_\perp^\prime \tau} q_{p_\perp \tau}}}{A} \bigl \lbrace
 \ii \tau \sin \vartheta \, F_0(\rho, \rho^\prime) {\bm e}_\varphi \pm
 \label{eq:polar-s-matrix} \\
 & \pm [-\tau \sin \vartheta \, G_0 (\rho, \rho^\prime)
 {\bm e}_r + \zeta_0 (\rho, \rho^\prime) {\bm e}_z] \bigr \rbrace, \nonumber
\end{align}
with upper (lower) sign corresponding to $m = 0$ ($\bar{1}$). This equation
reduces to \eqref{eq:mag-vortex} when $p_\perp = p^\prime_\perp$ (i.e.
$\rho^\prime = \rho$ and $F_0 = 0$).

\bibliographystyle{apsrev4-1}
\bibliography{references}
\end{document}